\documentclass[twocolumn]{aastex63}

\usepackage{xcolor}
\received{\today}
\revised{}
\accepted{\date{09/14/19}}

\submitjournal{AJ}

\shorttitle{MUSEpack}
\shortauthors{Zeidler et al.}

\NewPageAfterKeywords
\begin{document}

\title{The young massive star cluster Westerlund 2 observed with MUSE. \\
	II. MUSEpack -- a Python package to analyze the kinematics of young star clusters}

\correspondingauthor{Peter Zeidler}
\email{zeidler@stsci.edu}

\author[0000-0002-6091-7924]{Peter Zeidler}
\affil{Department of Physics and Astronomy, Johns Hopkins University, Baltimore, MD 21218, USA}
\affil{Space Telescope Science Institute, 3700 San Martin Drive, Baltimore, MD 21218, USA}

\author{Antonella Nota}
\affil{Space Telescope Science Institute, 3700 San Martin Drive, Baltimore, MD 21218, USA}
\affil{ESA, SRE Operations Devision, Spain}

\author[0000-0003-2954-7643]{Elena Sabbi}
\affil{Space Telescope Science Institute, 3700 San Martin Drive, Baltimore, MD 21218, USA}

\author{Peter Luljak}
\affil{Harvey Mudd College, 301 Platt Boulevard, Claremont, CA 91711, USA}
\affil{Space Telescope Science Institute, 3700 San Martin Drive, Baltimore, MD 21218, USA}

\author[0000-0002-5456-523X]{Anna F. McLeod}
\affil{Department of Astronomy, University of California Berkeley, Berkeley, CA 94720, USA} 
\affil{Department of Physics and Astronomy, Texas Tech University, PO Box 41051, Lubbock, TX 79409, USA}

\author{Eva K. Grebel}
\affil{Astronomisches Rechen-Institut, Zentrum f\"ur Astronomie der Universit\"at Heidelberg, M\"onchhofstra{\ss}e 12--14, D-68120 Heidelberg, Germany} 

\author[0000-0001-5171-5629]{Anna Pasquali}
\affil{Astronomisches Rechen-Institut, Zentrum f\"ur Astronomie der Universit\"at Heidelberg, M\"onchhofstra{\ss}e 12--14, D-68120 Heidelberg, Germany} 

\author[0000-0002-0986-4759]{Monica Tosi}
\affil{INAF--Osservatorio di Astrofisica e Scienza dello Spazio di Bologna, Via Gobetti 93/3, I-40129 Bologna, Italy}

\begin{abstract}

We mapped the Galactic young massive star cluster Westerlund 2 (Wd2) with the integral field spectrograph MUSE (spatial resolution: $0.2\,{\rm arcsec}\,{\rm px}^{-1}$, spectral resolution: $\Delta \lambda = 1.25 {\rm \AA}$, wavelength range 4600--$9350 {\rm \AA}$) mounted on the VLT, as part of an on-going study to measure the stellar and gas kinematics of the cluster region. In this paper we present the fully reduced dataset and introduce our new Python package ``MUSEpack", which we developed to measure stellar radial velocities with an absolute precision of 1--$2\,{\rm km}\,{\rm s}^{-1}$ without the necessity of a spectral template library. This novel method uses the two-dimensional spectra and an atomic transition line library to create templates around strong absorption lines for each individual star. The code runs fully automatically on multi-core machines, which makes it possible to efficiently determine stellar radial velocities of a large number of stars with the necessary precision to measure the velocity dispersion of young star clusters. MUSEpack also provides an enhanced method for removing telluric lines in crowded fields without sky exposures and a Python wrapper for ESO's data reduction pipeline.

We observed Wd2 with a total of 11 short and 5 long exposures to cover the bright nebular emission and OB stars, as well as the fainter pre-main sequence stars down to $\sim 1\,{\rm M}_\odot$. The survey covers an area of $\sim 11\,{\rm arcmin}^2$ ($15.8\,{\rm pc}^2$). In total, we extracted 1,725 stellar spectra with a mean S/N$>5$ per pixel. A typical radial velocity (RV) uncertainty of $4.78\,{\rm km}\,{\rm s}^{-1}$, $2.92\,{\rm km}\,{\rm s}^{-1}$, and $1.1\,{\rm km}\,{\rm s}^{-1}$ is reached for stars with a mean S/N$>10$, S/N$>20$, S/N$>50$ per pixel, respectively. Depending on the number of spectral lines used to measure the RVs, it is possible to reach RV accuracies of $0.9\,{\rm km}\,{\rm s}^{-1}$, $1.3\,{\rm km}\,{\rm s}^{-1}$, and $2.2\,{\rm km}\,{\rm s}^{-1}$ with $\geq 5$, 3--4, and 1--2 spectral lines, respectively. The combined statistical uncertainty on the radial velocity measurements is $1.10\,{\rm km}\,{\rm s}^{-1}$.

\end{abstract}


\section{Introduction}
\label{sec:introduction}

The majority of stars are born in clustered environments through the collapse of giant molecular clouds \citep[GMCs, e.g., ][]{Lada_03}. The initial mass and density of the GMC defines the number of stars formed in the newly born star cluster and its stellar density. The stellar densities range from low-density systems in the Milky Way (MW), such as the Upper Scorpius OB association \citep[$\le 0.1\,{\rm M}_\odot\,{\rm pc}^{-3}$, e.g., ][]{Preibisch_08}, via loosely bound open clusters, to more massive systems, like Westerlund~2 \citep[$\sim 6.6\times 10^3\,{\rm M}_\odot\,{\rm pc}^{-3}$, e.g.,][hereafter Wd2]{Westerlund_61, Zeidler_15,Zeidler_16b,Zeidler_17,Zeidler_18}, NGC~346 \citep[$\sim 64\,{\rm M}_\odot\,{\rm pc}^{-3}$, e.g.,][]{Sabbi_08}, or NGC~3603 \citep[$\sim 1.5 \times 10^4\,{\rm M}_\odot\,{\rm pc}^{-3}$, e.g.,][]{Sung_04,Stolte_04,Rochau_10,Pang_11,Pang_13} to the most massive young superstar clusters, like R136 \citep[4.8--$24\times10^4\,{\rm M}_\odot\,{\rm pc}^{-3}$, e.g.,][]{Hunter_95,Crowther_16} in the Large Magellanic Cloud (LMC). Despite numerous studies of the MW and Magellanic Cloud young star clusters, their formation processes and the cluster evolution in the first few Myr are not well understood. On the one hand, we observe star clusters that appear to have formed in one central star burst (monolithic cluster formation e.g., \citealt{Lada_84a, Bastian_06}), as was suggested for the central star cluster HD97950 in NGC~3603 \citep{Banerjee_15a}. We also observe hierarchical cluster formation, in which a star cluster forms via subsequent merging of smaller sub-clusters \citep{Larson_81, Longmore_14, Dale_15a}, which is seen for many MW and Magellanic Cloud star clusters (e.g., Wd2, \citealt{Zeidler_15}; NGC~346, \citealt{Sabbi_07a,Sabbi_08}; or in multiple star clusters of the \ion{H}{2} region 30~Dor, \citealt{Sabbi_12,Sabbi_13,Sabbi_16}). The subsequent evolution of a newly born star cluster is regulated not only by the stellar density and total mass (the stellar mass function (MF)), but also by stellar kinematics.

While it is possible to measure stellar proper motions using long-baseline observations from ground and space, like multi-epoch \textit{Hubble} Space Telescope observations \citep[HST, e.g.,][]{Sabbi_19} or the Gaia satellite \citep{Gaia_16, Gaia_18}, the development of wide-field integral field units (IFUs) has made it feasible to spectroscopically map resolved, nearby star forming regions and young star clusters to also measure the radial velocities (RVs) with a similar telescope-time efficiency providing 3D stellar kinematics. 

The large field of view (FOV) of $1~{\rm arcmin}^2$ of the Multi Unit Spectrographic Explorer \citep[MUSE,][]{Bacon_10} mounted in the Nasmyth focus of UT4 at the Very Large Telescope (VLT) at the European Southern Observatory (ESO), Chile, has proven to be very effective in studying stellar RVs and gas kinematics, including feedback processes in star forming regions and young star clusters \citep[e.g.,][]{McLeod_15,McLeod_18,Zeidler_18}. Stellar RVs can be measured to an accuracy of $\sim 1\,{\rm km}\,{\rm s}^{-1}$ by cross-correlating the extracted spectra with spectral templates \citep{Zeidler_18, Kamann_13,Kamann_16,Kamann_18}. The spatial sampling of MUSE is $0.2\,{\rm arcsec}\,{\rm px}^{-1}$ with a spectral resolution of $\Delta \lambda = 1.25 {\rm \AA}$ observing at optical wavelengths between 4600 and $9350 {\rm \AA}$.

For young star clusters, where the majority of stars are still in their pre-main-sequence phase no proper optical spectral template libraries exist. Our new method, presented in this work, is independent of spectral template libraries. We show that our Python package \dataset[MUSEpack]{\doi{10.5281/zenodo.3433996}} is able to measure RVs with a $\sim 1\,{\rm km}\,{\rm s}^{-1}$ accuracy when applied to our VLT/MUSE, medium-resolution spectra of Wd2. The package has been developed specifically for MUSE data but can be used for any kind of IFU dataset.

This paper is structured as follows: In Sect.~\ref{sec:wd2} we introduce the work that has been done on the Galactic young star cluster Wd2. Sect.~\ref{sec:data} gives an overview of the MUSE dataset on Wd2. In Sect.~\ref{sec:musepack} we provide an overview of \dataset[MUSEpack]{\doi{10.5281/zenodo.3433996}}, while Sect.~\ref{sec:datareduction} provides a detailed description of the data reduction process. Sect.~\ref{sec:RVs} gives an introduction to stellar RV fitting and the performance of \dataset[MUSEpack]{\doi{10.5281/zenodo.3433996}}. In Sec.~\ref{sec:catalog} we introduce the created stellar RV catalog of Wd2. In Sect.~\ref{sec:summary} we summarize and conclude the work.

\section{Westerlund 2 - a young massive star cluster}
\label{sec:wd2}

\begin{figure*}[htb]
	\plotone{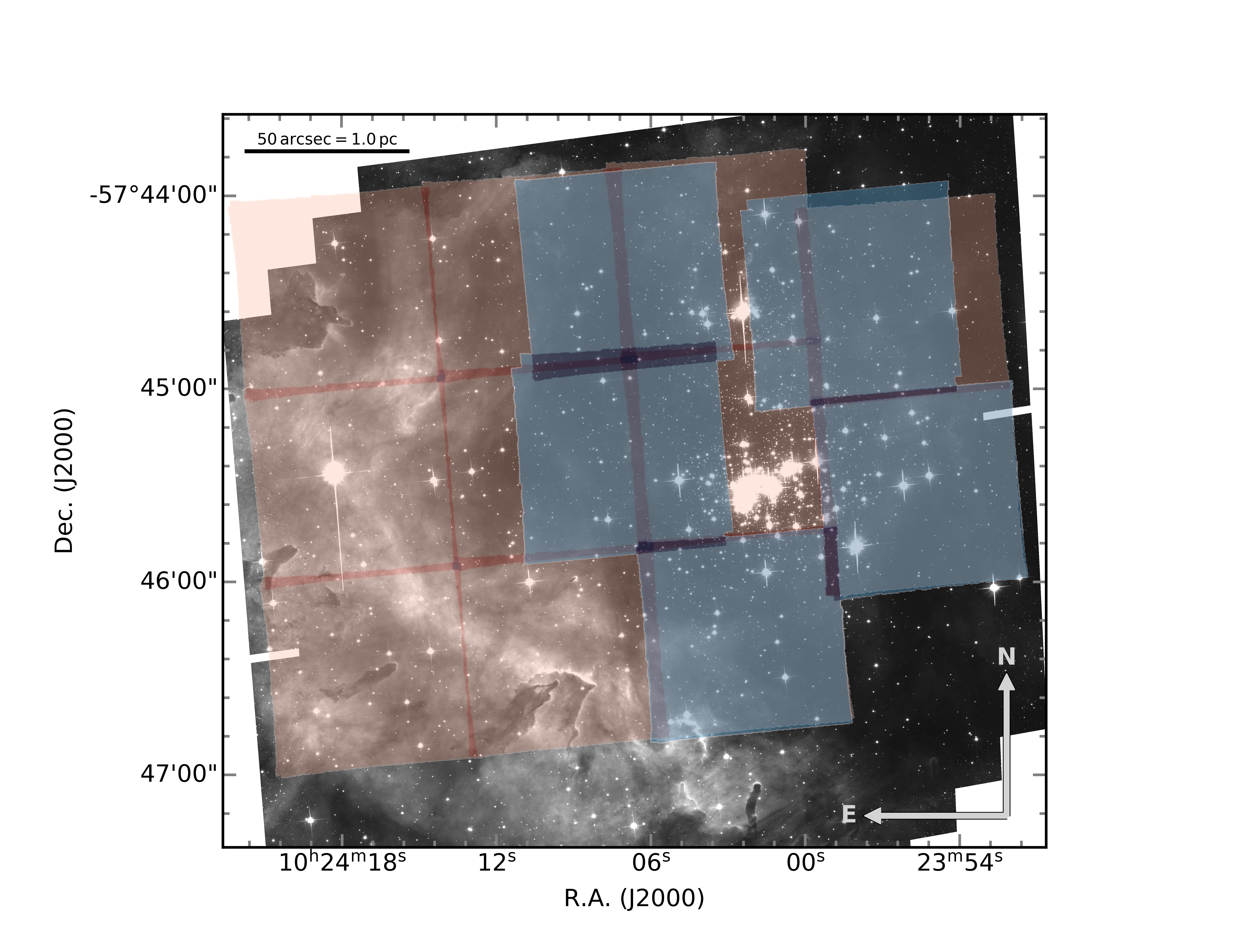}
	\caption{Coverage of the MUSE data plotted over the HST $F814W$ image of Wd2. The blue fields show the short exposures, while the red fields show the long exposures. In total we covered $\sim 11\,{\rm arcmin}^2$, which corresponds to $7.6\,{\rm pc}^2$ at the distance of Wd2.}
	\label{fig:coverage}
\end{figure*}

We are conducting a photometric and spectroscopic survey of the young massive star cluster (YMC) Wd2 \citep{Westerlund_61} in the \ion{H}{2} region RCW49 \citep{Rodgers_60} using 16 MUSE pointings (097.C-0044(A), 099.C-0248(A), P.I.: P.~Zeidler; see Fig.~\ref{fig:coverage}) with two different exposure times to obtain stellar RVs down to $\sim 1\,{\rm M}_\odot$ in combination with multi-band optical and near-infrared HST observations (ID: 13038, PI: A. Nota). Wd2 is a Galactic YMC located in the Sagittarius spiral arm, at a distance of $\sim4$\,kpc and an age of $<2$\,Myr consisting of two coeval sub-clumps, namely the Main Cluster and the Northern Clump \citep{Vargas_Alvarez_13,Hur_15,Zeidler_15}. We derived the stellar MF \citep{Zeidler_17} and showed that Wd2 is highly mass-segregated, which is, given its young age, likely primordial. In \citet{Zeidler_18}, we presented six MUSE pointings of the cluster center that were observed in the first (097.C-0044(A)) of two observing runs. Using these data we showed that 1) it is possible to obtain stellar and gas RVs with an accuracy of $\sim 1\,{\rm km}\,{\rm s}^{-1}$ in nearby YMCs, 2) we can measure spectral types of main and pre-main-sequence stars, and 3) Wd2 shows a bimodal velocity structure with two peaks at $\left(8.10 \pm 1.53 \right) \,{\rm km}\,{\rm s}^{-1}$ and $\left(25.41 \pm 3.15 \right)\,{\rm km}\,{\rm s}^{-1}$. The RV dispersion of both peaks is in agreement with the RV dispersion seen in other YMCs such as Trumpler~14 \citep{Kimiki_18}, the Carina Nebula \cite[e.g.,][]{Smith_08b}, or NGC~3603 \citep{Rochau_10,Pang_13}. Because of the limited data of only 72 stars, we could not properly address the physical origin of such bimodal RV distribution.

\section{The dataset}
\label{sec:data}

We observed Wd2 with MUSE in extended mode\footnote{MUSE can observe with two spectral ranges, the nominal mode (4800--9350\,\AA) and the extended mode (4600--9350\,\AA).}, both with and without adaptive optics (AO) during the ESO periods 97 and 99 (Program ID: 097.C-0044(A), 099.C-0248(A), PI: P.~Zeidler). In total, we acquired 16 pointings with 3 dither positions each, with a total telescope time (exposure time + overheads) of almost 21\,h. The dither pattern follows a 90 and 180 degrees rotation strategy to minimize detector defects and impurities. Eleven out of the sixteen pointings are obtained with a total exposure time of 660\,s (hereafter: short exposures), while the remaining five pointings have a total exposure time of 10,800\,s (hereafter: long exposures). An overview of all pointings is given in Tab.~\ref{tab:dataoverview}.

The two different exposure times were chosen to cover a large dynamic range. The short exposures (blue fields in Fig.~\ref{fig:coverage}) cover the complete cluster area including the gas rim of the surrounding \ion{H}{2} region to study the massive stars and the gas, while the long exposures (red fields in Fig.~\ref{fig:coverage}) are designed to study the fainter pre-main-sequence population. The long exposures cover the immediate vicinity of the cluster core avoiding the central regions and the gas, in order to avoid saturating a major fraction of the data cubes.

Four out of the five long exposures are observed using the wide-field AO mode of MUSE. This mode uses the four laser guide stars for ground-layer correction of the atmosphere's distortions resulting in a improvement of the seeing by a factor of two. Because the lasers are Na-lasers, a notch filter must be used to block the light between 5780 and 5990\AA, which results in a gap in the spectrum\footnote{For detailed information we refer to the VLT/MUSE handbook \url{https://www.eso.org/sci/facilities/paranal/instruments/muse/doc.html}.}.

\begin{deluxetable*}{cccccrccc}[htb]
	\tablecaption{Overview of the observations \label{tab:dataoverview}}
	\tabletypesize{\scriptsize}
	\tablecolumns{9}
	\tablewidth{0pt}
	\tablehead{
	\multicolumn{2}{c}{Coordinates} & \colhead{Date} & \colhead{OB-Name }& \colhead{OB-ID} & \colhead{$t_{\rm exp}$} &  \colhead{Seeing} & \colhead{Airmass}  & \colhead{AO}\\
		\colhead{R.A.} & \colhead{Dec.} & \colhead{(YYYY-mm-dd)} & \colhead{}& \colhead{} & \colhead{(s)} & \colhead{(arcsec)} & \colhead{} &  \colhead{} 
	}
	\startdata
	10:24:02.25 &-57:46:17.36 & 2018-03-06 & LONG\_1a & 1661895 & 3600 & 0.41-0.43 & 1.255-1.370 & no \\
	10:24:02.25 &-57:46:17.36 & 2017-12-21 & LONG\_1b & 1661899 & 3600 & 0.71-0.58 & 1.251-1.201 & no \\
	10:24:02.25 &-57:46:17.36 & 2018-03-06 & LONG\_1c & 1661903 & 3600 & 0.54-0.54 & 1.383-1.593 & no \\[3pt]
	10:24:07.17 &-57:45:22.75 & 2018-03-10 & LONG\_2a\_AO & 1987342 & 3600 & 0.94-0.51 & 1.466-1.309 & yes \\
	10:24:07.17 &-57:45:22.75 & 2018-03-11 & LONG\_2b\_AO & 1987346 & 3600 & 1.00-0.96 & 1.282-1.212 & yes \\
	10:24:07.17 &-57:45:22.75 & 2018-03-11 & LONG\_2c\_AO & 1987350 & 3600 & 0.53-0.49 & 1.206-1.197 & yes \\[3pt]
	10:23:55.47 &-57:45:31.39 & 2018-03-12 & LONG\_3a\_AO & 1987371 & 3600 & 0.59-0.46 & 1.197-1.236 & yes \\
	10:23:55.47 &-57:45:31.39 & 2018-03-11 & LONG\_3b\_AO & 1987367 & 3600 & 0.44-0.57 & 1.200-1.249 & yes \\
	10:23:55.47 &-57:45:31.39 & 2018-03-12 & LONG\_3c AO & 1987375 & 3600 & 0.59-0.58 & 1.279-1.413 & yes \\[3pt]
	10:23:58.21 &-57:44:34.58 & 2018-03-13 & LONG\_4a\_AO & 1987380 & 3600 & 0.43-0.47 & 1.468-1.310 & yes \\
	10:23:58.21 &-57:44:34.58 & 2018-03-13 & LONG\_4b\_AO & 1987384 & 3600 & 0.54-0.43 & 1.294-1.217 & yes \\
	10:23:58.21 &-57:44:34.58 & 2018-03-13 & LONG\_4c\_AO & 1987388 & 3600 & 0.57-0.42 & 1.211-1.196 & yes \\[3pt]
	10:24:06.89 &-57:44:24.14 & 2018-03-14 & LONG\_5a\_AO & 1987393 & 3600 & 0.41-0.57 & 1.241-1.198 & yes \\
	10:24:06.89 &-57:44:24.14 & 2018-03-14 & LONG\_5b\_AO & 1987397 & 3600 & 0.88-0.64 & 1.195-1.211 & yes \\
	10:24:06.89 &-57:44:24.14 & 2018-03-20 & LONG\_5c\_AO & 1987401 & 3600 & 0.67-0.57 & 1.335-1.237 & yes \\[3pt]
	10:23:56.11 &-57:44:33.14 & 2018-01-21 & SHORT-1 & 1661920 & 220 & 1.26-1.50 & 1.234-1.240 & no \\
	10:23:56.11 &-57:44:33.14 & 2018-01-21 & SHORT-1 & 1661920 & 220 & 1.47-1.30 & 1.247-1.253 & no \\
	10:23:56.11 &-57:44:33.14 & 2018-01-21 & SHORT-1 & 1661920 & 220 & 1.50-1.41 & 1.240-1.246 & no \\[3pt]
	10:23:55.48 &-57:45:31.43 & 2018-01-21 & SHORT-1 & 1661920 & 220 & 1.30-1.24 & 1.254-1.260 & no \\
	10:23:55.48 &-57:45:31.43 & 2018-01-21 & SHORT-1 & 1661920 & 220 & 1.24-1.21 & 1.261-1.268 & no \\
	10:23:55.48 &-57:45:31.43 & 2018-01-21 & SHORT-1 & 1661920 & 220 & 1.35-1.25 & 1.269-1.276 & no \\[3pt]
	10:24:17.45 &-57:45:29.30 & 2018-01-25 & SHORT-2 & 1661962 & 220 & 0.52-0.62 & 1.310-1.320 & no \\
	10:24:17.45 &-57:45:29.30 & 2018-01-25 & SHORT-2 & 1661962 & 220 & 0.51-0.52 & 1.300-1.309 & no \\
	10:24:17.45 &-57:45:29.30 & 2018-01-25 & SHORT-2 & 1661962 & 220 & 0.52-0.62 & 1.310-1.320 & no \\[3pt]
	10:24:16.83 &-57:46:27.62 & 2018-01-25 & SHORT-2 & 1661962 & 220 & 0.71-0.60 & 1.332-1.342 & no \\
	10:24:16.83 &-57:46:27.62 & 2018-01-25 & SHORT-2 & 1661962 & 220 & 0.60-0.51 & 1.343-1.354 & no \\
	10:24:16.83 &-57:46:27.62 & 2018-01-25 & SHORT-2 & 1661962 & 220 & 0.62-0.60 & 1.320-1.330 & no \\[3pt]
	10:24:18.08 &-57:44:30.98 & 2018-01-25 & SHORT-2 & 1661962 & 220 & 0.59-0.57 & 1.273-1.281 & no \\
	10:24:18.08 &-57:44:30.98 & 2018-01-25 & SHORT-2 & 1661962 & 220 & 0.57-0.49 & 1.281-1.290 & no \\
	10:24:18.08 &-57:44:30.98 & 2018-01-25 & SHORT-2 & 1661962 & 220 & 0.61-0.59 & 1.265-1.272 & no \\[3pt]
	10:24:02.27 &-57:46:17.40 & 2016-06-03 & SHORT-MID\_1 & 1323877 & 220 & 0.74-0.61 & 1.270-1.277 & no \\
	10:24:02.27 &-57:46:17.40 & 2016-06-02 & SHORT-MID\_1 & 1323877 & 220 & 0.93-0.67 & 1.260-1.266 & no \\
	10:24:02.27 &-57:46:17.40 & 2016-06-02 & SHORT-MID\_1 & 1323877 & 220 & 0.83-0.85 & 1.251-1.257 & no \\[3pt]
	10:24:02.90 &-57:45:19.12 & 2016-06-02 & SHORT-MID\_1 & 1323877 & 220 & 0.75-0.79 & 1.242-1.248 & no \\
	10:24:02.90 &-57:45:19.12 & 2016-06-02 & SHORT-MID\_1 & 1323877 & 220 & 0.78-0.85 & 1.235-1.240 & no \\
	10:24:02.90 &-57:45:19.12 & 2016-06-02 & SHORT-MID\_1 & 1323877 & 220 & 0.79-0.96 & 1.228-1.232 & no \\[3pt]
	10:24:03.53 &-57:44:20.80 & 2016-06-02 & SHORT-MID\_1 & 1323877 & 220 & 1.10-0.81 & 1.221-1.226 & no \\
	10:24:03.53 &-57:44:20.80 & 2016-06-02 & SHORT-MID\_1 & 1323877 & 220 & 0.82-0.67 & 1.211-1.214 & no \\
	10:24:03.53 &-57:44:20.80 & 2016-06-02 & SHORT-MID\_1 & 1323877 & 220 & 0.98-0.95 & 1.216-1.220 & no \\[3pt]
	10:24:09.55 &-57:46:22.51 & 2016-06-03 & SHORT-MID\_2 & 1323880 & 220 & 0.52-0.73 & 1.393-1.406 & no \\
	10:24:09.55 &-57:46:22.51 & 2016-06-03 & SHORT-MID\_2 & 1323880 & 220 & 0.53-0.63 & 1.412-1.426 & no \\
	10:24:09.55 &-57:46:22.51 & 2016-06-03 & SHORT-MID\_2 & 1323880 & 220 & 0.66-0.49 & 1.375-1.388 & no \\[3pt]
	10:24:10.18 &-57:45:24.19 & 2016-06-03 & SHORT-MID\_2 & 1323880 & 220 & 0.68-0.69 & 1.359-1.370 & no \\
	10:24:10.18 &-57:45:24.19 & 2016-06-03 & SHORT-MID\_2 & 1323880 & 220 & 0.70-0.65 & 1.343-1.354 & no \\
	10:24:10.18 &-57:45:24.19 & 2016-06-03 & SHORT-MID\_2 & 1323880 & 220 & 0.66-0.62 & 1.328-1.338 & no \\[3pt]
	10:24:10.81 &-57:44:25.91 & 2016-06-03 & SHORT-MID\_2 & 1323880 & 220 & 0.63-0.64 & 1.301-1.310 & no \\
	10:24:10.81 &-57:44:25.91 & 2016-06-03 & SHORT-MID\_2 & 1323880 & 220 & 0.71-0.75 & 1.314-1.324 & no \\
	10:24:10.81 &-57:44:25.91 & 2016-06-03 & SHORT-MID\_2 & 1323880 & 220 & 0.78-0.58 & 1.289-1.298 & no \\
	\enddata
	\tablecomments{Overview over the MUSE observations of Wd2 (Programm ID: 097.C-0044(A), 099.C-0248(A)). In Column 1 and 2 the coordinates are given. Column 3 shows the observation dates. Columns 4 and 5 indicate the name of each exposure and the observing block (OB) ID, respectively. Column 6 lists the exposure time. Column 7 indicates the seeing and Column 8 shows the airmass. Column 9 shows if this pointing was obtained using WFM-AO. OBs that had to be repeated do not appear in this table. We grouped the three exposures that are combined to one pointing.}
\end{deluxetable*}

\section{MUSEpack}
\label{sec:musepack}

To reduce and analyze our IFU dataset of the YMC Wd2, we developed \dataset[MUSEpack]{\doi{10.5281/zenodo.3433996}}. It is a Python-based package, which contains two main classes\footnote{A class is special construct for object-oriented programming languages that has the capability of bundling data and functionality together.}: \texttt{MUSEreduce} and \texttt{RV\_spectrum}. \dataset[MUSEpack]{\doi{10.5281/zenodo.3433996}} can be downloaded from  \url{https://github.com/pzeidler89/MUSEpack.git}. We will introduce the capabilities of the various classes and their modules throughout this paper. A full description of the package can be found on \url{https://musepack.readthedocs.io/en/latest/index.html}. There, we also provide some basic examples to show the features of \dataset[MUSEpack]{\doi{10.5281/zenodo.3433996}} and to provide the user with an easy start.

\subsection{MUSEreduce}

\texttt{MUSEreduce} is a Python wrapper\footnote{A Python script that calls functions in other programming languages.} for the standard reduction pipeline \citep{Weilbacher_12,Weilbacher_14}, based on ESO\footnote{\url{https://www.eso.org/sci/software/pipelines/muse/}} Reflex (ESORex), which provides a user-friendly \texttt{json} file\footnote{A \texttt{json} or JavaScript Object Notation file is an open-standard, human-readable text file format containing data objects as attribute-value pairs.} to set the input parameters for ESORex. Additionally, \texttt{MUSEreduce} is meant to sort the raw data and assign the appropriate calibration files to each science exposure. In contrast to ESO's graphical user interface (GUI)-based data manager GASGANO\footnote{\url{https://www.eso.org/sci/software/gasgano.html}}, \texttt{MUSEreduce} is Python-based, which makes it platform-independent and easy to use for remote execution via a secure shell (SSH) connection\footnote{An SSH connection is a cryptographic network protocol to connect network services securely over an unsecured network.}. \texttt{MUSEreduce} supports the three main observing modes of MUSE: the wide-field mode without adaptive optics (WFM-NOAO), the wide-field mode with adaptive optics (WFM-AO), and the narrow-field mode with adaptive optics (NFM-AO). For the quality assessment of the code we used data obtained with the three different observing modes. For the WFM-NOAO and WFM-AO we used data of the Galactic YMC Wd2 (097.C-0044(A), 099.C-0248(A), P.I.: P.~Zeidler) representative of a crowed field with a high dynamic range (\citet{Zeidler_18}, and this work). We also used data of the blue compact dwarf galaxy J0291+0721 (096.B-0212(A), P.I.: B. James), a faint extended object \citep{James_19}. The data of J0291+0721 was obtained including two sky fields. To test the NFM-AO we used observations of two T-Tauri disks (60.A-9482(A), P.I.: J. Girard), which were also obtained using sky fields \citep{Girard_19}.

\subsection{RV\_spectrum}

\texttt{RV\_spectrum} is the main workhorse routine to fit spectral lines to extracted stellar and gas spectra and to fit RVs with an accuracy of 1--2$\,{\rm km}\,{\rm s}^{-1}$. The novelty compared to other RV measurement techniques is that it does not require stellar spectral template libraries. To be able to use the full capabilities of this class the user should install the non-standard Python packages PampelMuse \citep{Kamann_13}, Penalized Pixel-Fitting \citep[pPXF][]{Cappellari_04,Cappellari_17}, and pyspeckit \citep{Ginsburg_11}.

\section{Data reduction}
\label{sec:datareduction}

The data were reduced using \texttt{MUSEreduce}, a Python class within \dataset[MUSEpack]{\doi{10.5281/zenodo.3433996}} (see Sect.~\ref{sec:musepack}) using ESO's standard reduction pipeline (v.2.4.2). When available, we used the reduced calibration files (MASTERFLAT, MASTERBIAS. WAVECAL, LSF, TWILIGHT) provided by ESO. If the reduced calibration files were not provided, we executed the respective data reduction steps to create them from the raw calibration files. All intermediate and final products were checked by eye in order to ensure that the data reduction processes and the alignment of the dither positions were executed correctly.

During these quality checks we discovered a Moir\'{e}-like pattern in the southern-most long exposure (LONG\_1). After a thorough analysis we concluded that it is most likely due to the sky conditions  under which the individual dither positions were obtained. While the exposures LONG\_1a and LONG\_1c were obtained with Moon condition ``bright", the exposure LONG\_1b was observed with Moon condition ``dark" (during a different night, see Tab.~\ref{tab:dataoverview}). We used a simulated lunar spectrum together with the sky spectrum of the exposure LONG\_1b to model the lunar contamination and to remove it from the data cube. This procedure is described in detail in Appendix~\ref{sec:moon}.

\subsection{Telluric line correction in crowded \ion{H}{2} regions}
\label{sec:telluric_correction}

To subtract the sky (dominated by telluric emission lines) from the science exposures, the MUSE data reduction pipeline uses a sky spectrum extracted from either a sky observation or the darkest regions of the science observation itself. This sky spectrum is convolved with the local, position-dependent line spread function (LSF) and together with a low-order continuum refitted to each spaxel. Subsequently, it is subtracted from the data cube.

For many observations, such as Galactic star clusters, obtaining sky fields is not feasible since the telescope has to be moved by several degrees to avoid observing the sky inside the parental gas and dust cloud. The high stellar crowding within these young star clusters and the gas and dust of the \ion{H}{2} region often does not leave enough ``empty" regions to properly determine a sky spectrum, yet a proper sky subtraction is necessary to remove contamination from the mostly dominant emission lines created by the Earth's atmosphere. Some of these telluric emission lines have the same wavelengths of typical \ion{H}{2} region emission lines, which are typically the Balmer lines, the oxygen lines, and atomic and molecular transition lines of oxygen, hydrogen, and some other trace elements. As a result, the extracted sky spectra overestimate the fluxes for these emission lines, which leads to an over-subtraction with a possible shift of the centroid.

To improve we altered the process how the data reduction pipeline handles the estimation of the sky spectrum under two main assumptions: 1) the fluxes of the \ion{H}{2} region emission lines are much stronger than the emitted fluxes from our atmosphere, and 2) the \ion{H}{2} region continuum is dominant over the telluric continuum. The sky spectrum is then extracted in the following manner:

\begin{itemize}
	\item[1] The data reduction pipeline performs a regular sky line and continuum extraction (\texttt{muse\_create\_sky}) for the darkest spaxels of each data cube creating the two files 'SKY\_LINES.fits' and 'SKY\_CONTINUUM.fits' containing the fitted sky line fluxes and the continuum, respectively.
	\item[2] The fluxes of all emission lines besides the \ion{OH}{0} and ${\rm O}_2$ lines is set to zero in the 'SKY\_LINES.fits' file.
	\item[3] The continuum flux is set to zero to avoid subtracting the continuum of the \ion{H}{2} region.
	\item[4] \texttt{muse\_create\_sky} is run again with the new 'SKY\_LINES.fits' and 'SKY\_CONTINUUM.fits' files as input. This is necessary to properly determine the fluxes of the \ion{OH}{0} and ${\rm O}_2$ emission lines that are blended with other emission lines (e.g. [\ion{O}{1}] $\lambda 5577$ and OH(1601)X-X(0701)Q11(03.5f) at $\lambda=5577.58273$\,\AA).
	\item[5] As last step the sky subtraction is handled in \texttt{muse\_scipost} by only applying the local LSF but not refitting the fluxes for each spaxel (skymethod = subtract-model).
\end{itemize}

\begin{figure*}[htb]
	\plotone{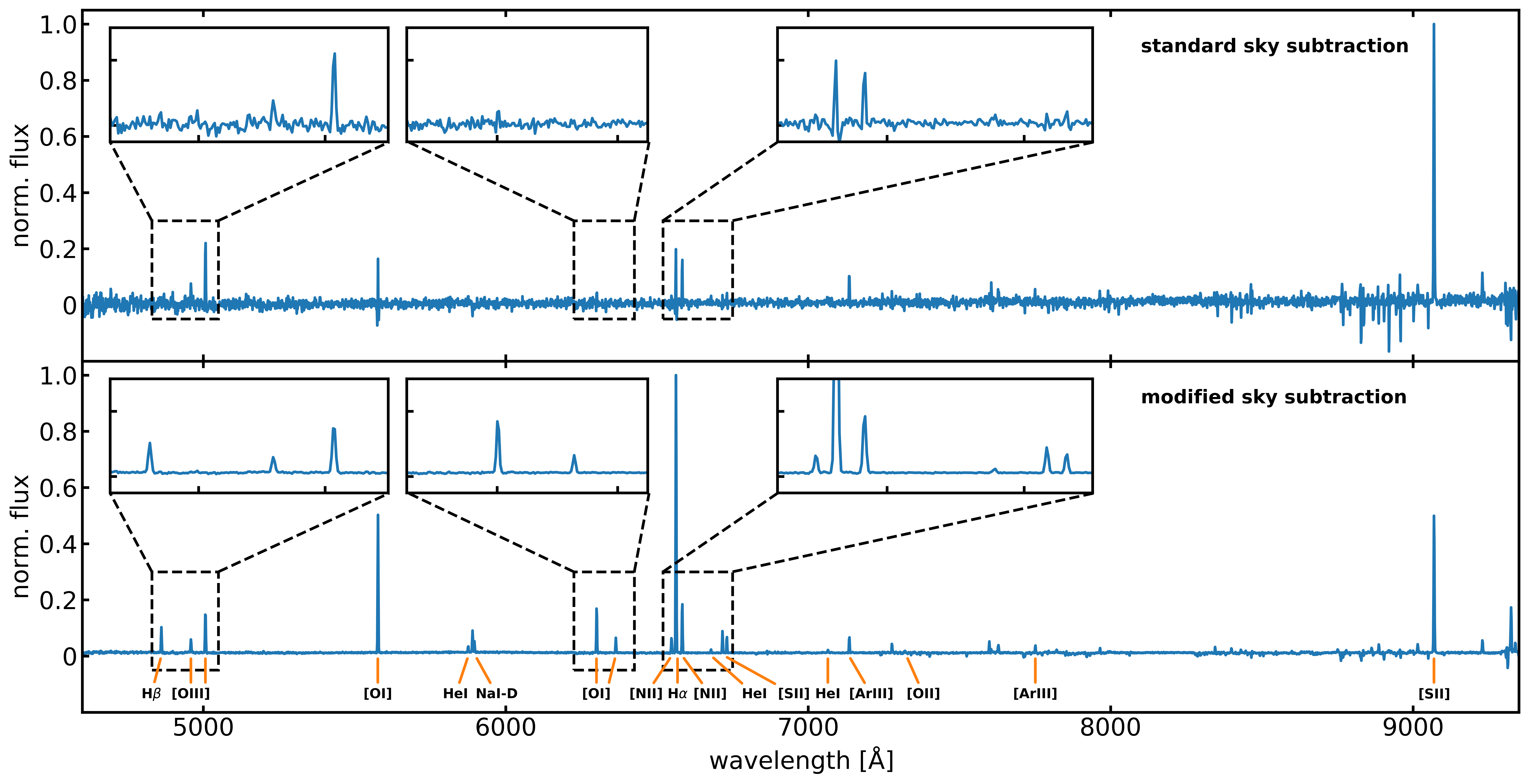}
	\caption{A sample spectrum extracted from a less luminous part of the \ion{H}{2} region. \textbf{Top:} The spectrum as it is extracted using the standard method provided by the MUSE data reduction pipeline. \textbf{Bottom:} The spectrum as it is extracted with our modified version, as described in Sect.~\ref{sec:telluric_correction}. Main nebular emission lines are marked. It becomes clear that many main nebular lines, such as [\ion{O}{1}]$\lambda5577$, H$\alpha$, H$\beta$, or the sodium doublet are properly recovered.}
	\label{fig:skysub}
\end{figure*}

To run this method we need to set the ``sky: modified" keyword in the config.json file to ``True". It is also important that the subtraction method is ``subtract-model" to avoid refitting the spectral lines to each spaxel. It is possible to properly recover many of the nebular emission lines, such as [\ion{O}{1}]$\lambda5577$, H$\alpha$ or H$\beta$ (see Fig.~\ref{fig:skysub}). For emission lines such as [\ion{O}{1}] $\lambda 5577$, the telluric contamination by blends can be significant, especially for long exposure times, which means that condition 1 is not satisfied anymore. Therefore, we recommend to use short exposure times to determine fluxes of the gas emission to minimize the residual telluric contamination in blends, and thus properly measure the flux of nebular emission lines. Stellar spectra are not affected by this since a local, star-by-star background subtraction is performed (see Sect.~\ref{sec:background}).

Tests have shown that if a sky field is used instead of the science exposure to obtain the sky spectrum, e.g., J0291+0721 (096.B-0212(A), P.I.: B. James), ESO's standard procedure for sky correction is recommended. The reason for this is that a refit of the fluxes of the sky emission lines to the science data may be necessary, caused by possible different observing conditions between the science and the sky exposures.

\subsection{Source detection and extraction}
\label{sec:surce_detection}
To extract the stellar spectra from the reduced MUSE data cubes we used the Python package PampelMuse \citep{Kamann_13}. PampelMuse performs a position- and-wavelength-dependent PSF fit for each layer of the MUSE data cube. As input a high-resolution photometric stellar source catalog must be provided\footnote{The catalog should be as complete as possible and go at least 2~mag deeper than the MUSE detection limit. Additionally, a much higher spatial resolution than MUSE is recommended to detect blends.}. This deep high-resolution catalog provides the necessary information to PampelMuse to de-blend the spectra of the majority of stars, especially in crowded regions, or to flag spectra accordingly where de-blending is not possible anymore. These spectra were not used in the further analysis. We used the HST multi-band photometric star catalog of the Wd2 region obtained by \citet{Zeidler_15}. When available, we used the \textit{F814W} photometry as brightness estimate, since this band roughly covers the wavelength range of the MUSE dataset, otherwise (e.g. for saturated sources) we use one of the other wide-band filters. For some very bright sources all filters were saturated. These stars were assigned  $F814W = 10$\,mag\footnote{This magnitude estimate is sufficient for the source detection and PampelMuse will properly determine the magnitudes based on the extracted spectra and the HST $F814W$ throughput curve.} to provide a catalog as complete as possible. To create an input catalog in the necessary PampelMuse format, the \dataset[MUSEpack]{\doi{10.5281/zenodo.3433996}} module \texttt{cubes.pampelmuse\_cat} can be used.
MUSE is a seeing limited instrument, which means that each point source covers multiple spaxels on the detector and each spectrum is extracted from multiple spaxels, which mitigates wavelength calibration uncertainties.

\subsection{Flux offsets between different pointings}
\label{sec:flux_offsets}
PampelMuse also calculates the stellar magnitudes by convolving the extracted stellar spectra with the throughput curve of the filter given in the source catalog. In the case of the Wd2 data the input magnitudes correspond to the Wide Field Camera (WFC) $F814W$ filter of the Advanced Camera for Surveys \citep[ACS,][]{ACS}. The ACS \textit{F814W} filter covers a wavelength range from $~6900\,{\rm \AA}$ to $9500\,{\rm \AA}$ while the longest wavelength of the MUSE spectra is $9350\,{\rm \AA}$. Therefore, the ACS $F814W$ covers about $150\,{\rm \AA}$ more of the stellar spectral energy distributions (SEDs) than the MUSE data and it is expected that the magnitudes extracted from the MUSE spectra are somewhat fainter than the ACS magnitudes. Additionally, the MUSE dataset was not observed under photometric conditions, which also introduces some flux uncertainties.

\begin{figure}[htb]
	\plotone{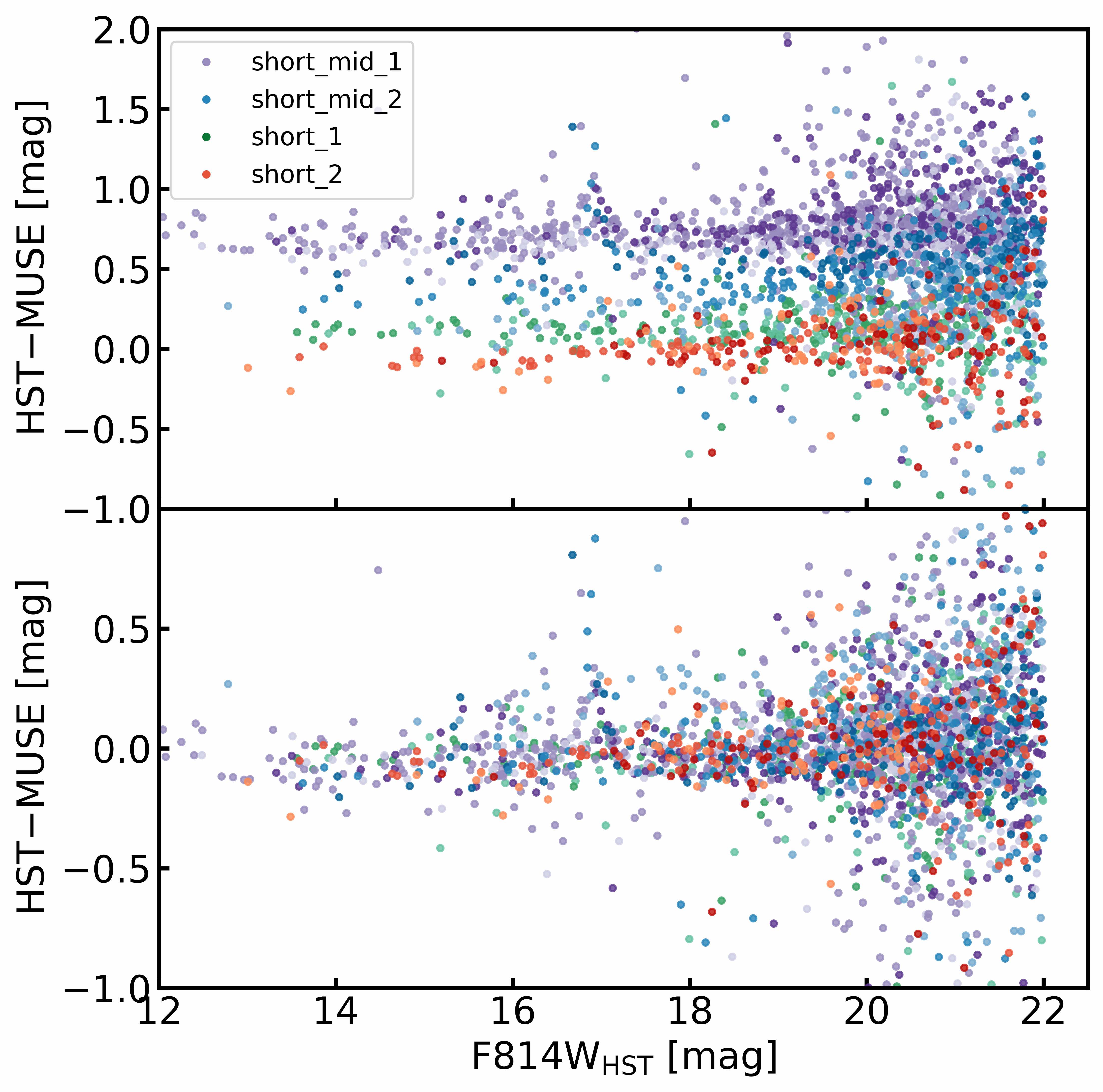}
	\caption{The offsets of the $F814W$ stellar magnitudes for the four short exposure OBs (top panel). The different shades for each color represent one pointing within the OB. In the bottom panel we show the corrected magnitudes.}
	\label{fig:mag_diff}
\end{figure}

To confirm the reliability of the MUSE data reduction we analyzed the stellar magnitudes and we find that the MUSE $F814W$ magnitudes are up to $\sim 0.7$\,mag brighter than the ACS $F814W$ magnitudes. Between different observing blocks (OBs) these offsets vary between 0\,mag and 0.77\,mag, which corresponds to a flux difference of up to a factor of two (see Fig.~\ref{fig:mag_diff}). We also measured the flux differences using sky emission to ensure that the offsets are not caused by stellar binary systems or blends. The variation between different pointings within one OB is typically small ($<0.1$\,mag). These small variations within an OB can be explained by changes in the observing conditions (clouds, temperature, etc.) since the observations were not taken under photometric conditions. The large offsets between different OBs are most likely caused by the calibration of the spectrophotometric standard star\footnote{The stars used for this calibration are automatically taken from the reference list of standard stars: \url{https://www.eso.org/sci/observing/tools/standards/spectra.html}.} including the time at which the standard star was observed in relation to the science observations, e.g., OB SHORT-MID\_1 was observed at the beginning of the night June 2/3 (22:48:04 -- 00:10:06 UT), while the corresponding standard star was observed at the end of the same night (09:52:36 -- 10:08:02 UT), partly into twilight (09:58:00 -- 11:11:00 UT).

To be able to compare fluxes between the various OBs of our dataset, we correct for these offsets in the following manner:

\begin{itemize}
	\item[1] We determine the median difference between the HST and MUSE $F814W$ magnitudes using a $3\sigma$-clipping to account for stellar variability and photometric uncertainties.
	\item[2] We transform the magnitude difference to a flux-scaling factor.
	\item[3] We correct each spaxel in the data cube for both the data and the variance cube.
	\item[4] We re-run CUBEFIT and GETSPECTRA of PampelMuse to obtain the spectra with the corrected fluxes.
\end{itemize}

The fluxes in the data cubes may be corrected using the \texttt{cubes.wcs\_cor} routine (see Sect.~\ref{sec:WCS} for more details) with ``correct\_flux=True". In the bottom panel of Fig~\ref{fig:mag_diff} we show the corrected magnitudes, whose differences are now close to zero. All long exposures, as well as the SHORT-1 and SHORT-2, show smaller variations (0.04 -- 0.14\,mag). These OBs were obtained in run 099.C-0248(A) where the observing conditions were generally better compared to run 097.C-0044(A), which was executed as a filler program.

One can say that the effective level of stellar crowding in the data is determined by the spectral type and magnitude of the stars and the depth of the data.

\subsection{The background subtraction}
\label{sec:background}

The extracted stellar spectra still show contamination from the emission of the local \ion{H}{2} region. Therefore, similarly to a photometric data reduction \citep[e.g., ][]{Zeidler_15} a local, star-by-star background subtraction has to be performed. Compared to photometry, this is a more challenging task for IFU observations. The PSF, the S/N ratio, and the stellar spectral energy distribution (SED) depend on wavelength. As a result, the background emission and the effective level of stellar crowding in the data, determined by the spectral type and magnitude of the stars, are a function of wavelength. PampelMuse estimates the background flux local to each star by dividing each spectral bin in equally-sized squares and weights the median flux in each square by the distance to the stars. The estimated background flux is subtracted from the stellar flux individually for each spectral layer. This method works well for regions with a low crowding or a background that is marginally changing within a few PSFs. Wd2 is a crowded cluster with a highly variable background (gas clouds) on small spatial scales. This leads to an over- or under-estimation of the background, especially at wavelengths where stellar absorption features overlap with nebular emission lines (e.g., Balmer lines or helium lines). We are currently working on a method with which we will be able to clean the spectra in a secondary step from any residual contamination. This will be published in a future paper and we will add the method to \dataset[MUSEpack]{\doi{10.5281/zenodo.3433996}}. For the further analysis in this work we are using the stars for which we extracted well-background-subtracted spectra.

\subsection{The extracted stars}
\label{sec:extracted_stars}
To extract the stellar spectra we decided to use a lower S/N limit of 5. Throughout the paper the S/N of the spectra always refers to the mean S/N per pixel. In total, we extracted 924 stellar spectra from the short exposures and 1,381 stellar spectra from the long exposures. Combining the two catalogs, we analyzed 1,726 stars with S/N$\ge 5$. In Tab.~\ref{tab:extr_stars} we provide an overview of the number of detected stars in the short and long exposures, as well as in the combined catalog. If a star was detected in both the long and short exposure, we use the spectrum with the higher S/N. For completeness we also provide the number of detected stars with S/N$<5$, which are not considered in any further analyses. One has to keep in mind that the long exposures (5 in total) only cover $\sim 45$\% of the area in comparison to the short exposures (11 in total). Additionally, the long exposures do not cover the cluster core.

\begin{deluxetable}{crrr}[htb]
	\tablecaption{The extracted stellar spectra \label{tab:extr_stars}}
	\tablecolumns{4}
	\tablehead{
		\colhead{S/N} & \colhead{short exp} & \colhead{long exp } & \colhead{unique }
	}
	\startdata
	$>50$  & 187 & 193 & 259 \\
	20--50  & 283 & 300 &  423\\
	10--20  & 192 & 359 &  427\\
	5--10  & 263 & 529 &  616\\
	$\ge 5$  & 925 & 1,381 &  1,725\\
	$<5$  & 2,656 & 2,395 &  3,318\\
	\enddata
	\tablecomments{The number of stellar spectra we extracted from the short (Column 2) and long exposures (Column 3) for different S/N ranges. For completeness we also provide the number of stars with a S/N$<5$, which are not considered in the further analyses. Column 4 contains the total number of detected stars. If a star was detected in both the long and short exposure, we use the spectrum with the higher S/N.}
\end{deluxetable}

\subsection{The world coordinate system - creating the mosaic}
\label{sec:WCS}

While pointing and guiding of the VLT are excellent (also for exposure times of 1\,h), the absolute world coordinate system (WCS) is less accurate leading to coordinate offsets between individual exposures. This is corrected by the MUSE data reduction pipeline when combining individual dither positions but not between individual OBs. To properly create mosaics and compare the data to other observations (e.g., HST, \textit{Spitzer}), a common WCS is needed. We decided to adopt the Gaia DR2 WCS. Therefore, as the first step we matched our photometric HST catalog \citep{Zeidler_15} to the Gaia DR2 catalog \citep{Gaia_16,Gaia_18} using the software CataPack\footnote{\url{http://www.bo.astro.it/\%7Epaolo/Main/CataPack.html}}. CataPack uses a 2D histogram technique to match multiple catalogs and to determine the coordinate transformation including a polynomial of degree n (defined by the user) for the coordinate transformation. The latter is then subsequently used to correct the coordinates of the HST catalog and the corresponding headers of the Wd2 MUSE images. Although the accuracy of the Gaia DR2 catalog is still limited in crowded regions, the stellar density in the outskirts of Wd2 is low enough for an accurate determination of the WCS. In total we used 1239 stars with a median uncertainty for the positions of the stars of $0.345$\,mas and $0.319$\,mas in R.A. and Dec., respectively.

PampelMuse determines the shift, rotation, and scale between the MUSE data cubes and the photometric input catalog. We use the PampelMuse solution to obtain a uniform transformation between the photometric stellar catalog and each individual data cube. This transformation is added as a new, primary WCS to the fits headers of the data cubes. For these transformations we used the \texttt{cubes.wcs\_cor} module from \dataset[MUSEpack]{\doi{10.5281/zenodo.3433996}}.

\begin{figure*}[htb]
	\plotone{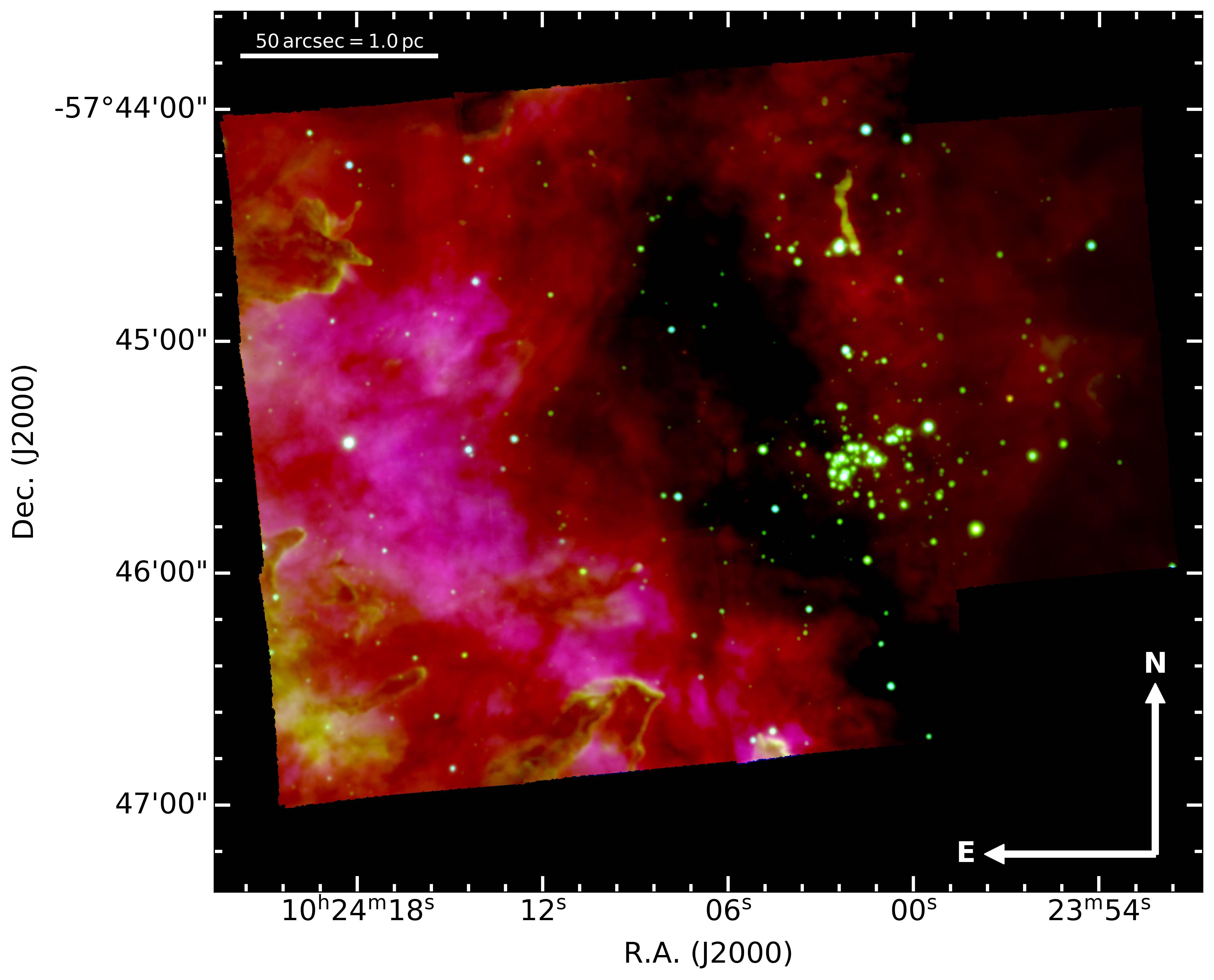}
	\caption{Mosaic of the MUSE dataset of Wd2 (short exposures only). To create this image the $H\alpha$ (red), [\ion{N}{2}]$\lambda 6583$\,\AA~(green), and [\ion{O}{3}]$\lambda 5007$\,\AA~(blue) nebular lines were used.}
	\label{fig:MUSE_RGB}
\end{figure*}

With this common WCS it is possible to properly mosaic layers of the individual cubes. The modules \texttt{cubes.linemaps} and \texttt{cubes.mosaics} provide the necessary processing. In Fig.~\ref{fig:MUSE_RGB} we show the RGB mosaic representing the short exposures using the slices for the H$\alpha$ (red), [\ion{N}{2}]$\lambda 6583$\,\AA~(green), and [\ion{O}{3}]$\lambda 5007$\,\AA~(blue) spectral lines.

\section{The radial velocities}
\label{sec:RVs}
Measuring accurate RVs in YMCs is challenging as the typical velocity dispersion in these clusters is on the order of $5\,{\rm km}/{\rm s}$. Traditionally, it has been necessary to obtain high resolution stellar spectra with fiber, slit, or echelle spectrographs. The development of large IFUs allows, for the first time, to map these resolved star clusters spectroscopically with a similar efficiency as photometric surveys. However, their typically low spectral resolution presents a challenge. Nevertheless, \citet{Kamann_13,Kamann_16,Kamann_18} showed for Galactic globular cluster (GC) stars that using MUSE spectra and a spectral library allows one to determine the RVs of stars to a 1--$2\,{\rm km}/{\rm s}$ accuracy. This method allowed them to study the rotation profile of these old star clusters but because no extensive optical pre-main sequence spectral libraries exist, applying this to young clusters is challenging

The novelty of our method is to create a template library based on the observed spectra, as well as a spectral line library. An introduction to this method was presented in \citet{Zeidler_18}. It uses strong stellar spectral lines in combination with a Monte-Carlo (MC) approach to measure the RVs. If not stated otherwise, from now on we use wavelengths in \AA ngstrom and the rest-frame for air, since this is the standard for the reduced MUSE data cubes, although \dataset[MUSEpack]{\doi{10.5281/zenodo.3433996}} does not depend on a specific rest-frame.

\subsection{The spectral templates}
\label{sec:templates}
To create stellar spectral templates we use the information provided by the Atomic Spectra Database of the National Institute of Standards and Technology (NIST\footnote{\url{https://www.nist.gov/pml/atomic-spectra-database}}). The templates are created around strong stellar absorption lines, such as \ion{Mg}{1} $\lambda\lambda$5367, 5172, 5183, \ion{Na}{1} $\lambda\lambda$5889, 5895, and \ion{Ca}{2} $\lambda\lambda$8498, 8542, 8662 for stars with a cool photosphere and \ion{He}{1} and \ion{He}{2} lines for stars with a hotter photosphere. We are not using the Balmer lines because the majority of stars are still in their pre-main-sequence phase and as a result these lines may be affected by accretion processes \citep[e.g.,][]{Zeidler_16b}.

\subsubsection{The Method}
The templates are created using the \texttt{line\_fitting} instance of the \dataset[MUSEpack]{\doi{10.5281/zenodo.3433996}} class \texttt{RV\_spectrum}, which uses pyspeckit \citep{Ginsburg_11} as core fitting routine. The spectral resolution of the MUSE data requires that \texttt{line\_fitting} properly handles blended lines. For a proper velocity fit it is necessary that the line of interest (primary line) is stronger than the blend in order to properly fit the centroid. To create a spectral template the following steps must be executed:

\begin{itemize}
	\item[1)] The creation of a spectrum instance using the \texttt{RV\_spectrum} class. This instance includes the wavelength, flux, and flux uncertainty array as well as all methods needed to run the RV fit.
	\item[2)] The reading of a catalog (\texttt{RV\_spectrum.catalog}) of lines that shall be fitted and used to measure RVs. This catalog must contain: the unique line name, the rest-frame wavelength of the spectral line, the start and end point of the fitting range, and the initial order of the polynomial used to fit the continuum. The order of the polynomial may be increased subsequently while fitting the spectra.
	\item[3)] The execution of the main fitting module (\texttt{line\_fitting}), which attempts to fit the spectral line profiles \citep[using a Voigt-profile, see Sect.~5.1 of][]{Zeidler_18} with a common continuum, taking into account wavelength limits, blends, and the instrument dispersion. The \texttt{line\_fitting} module is capable of running on multiple cores (one core per primary line). After the fit has converged, a fitted spectrum as well as the template are created and added to the instance. To create the template, \texttt{line\_fitting} uses the continuum and the fitted line parameters ($A$, $\sigma$, $\gamma$) but the rest-frame wavelengths from the spectral line catalog.
\end{itemize}

\begin{figure*}[htb]
	\plotone{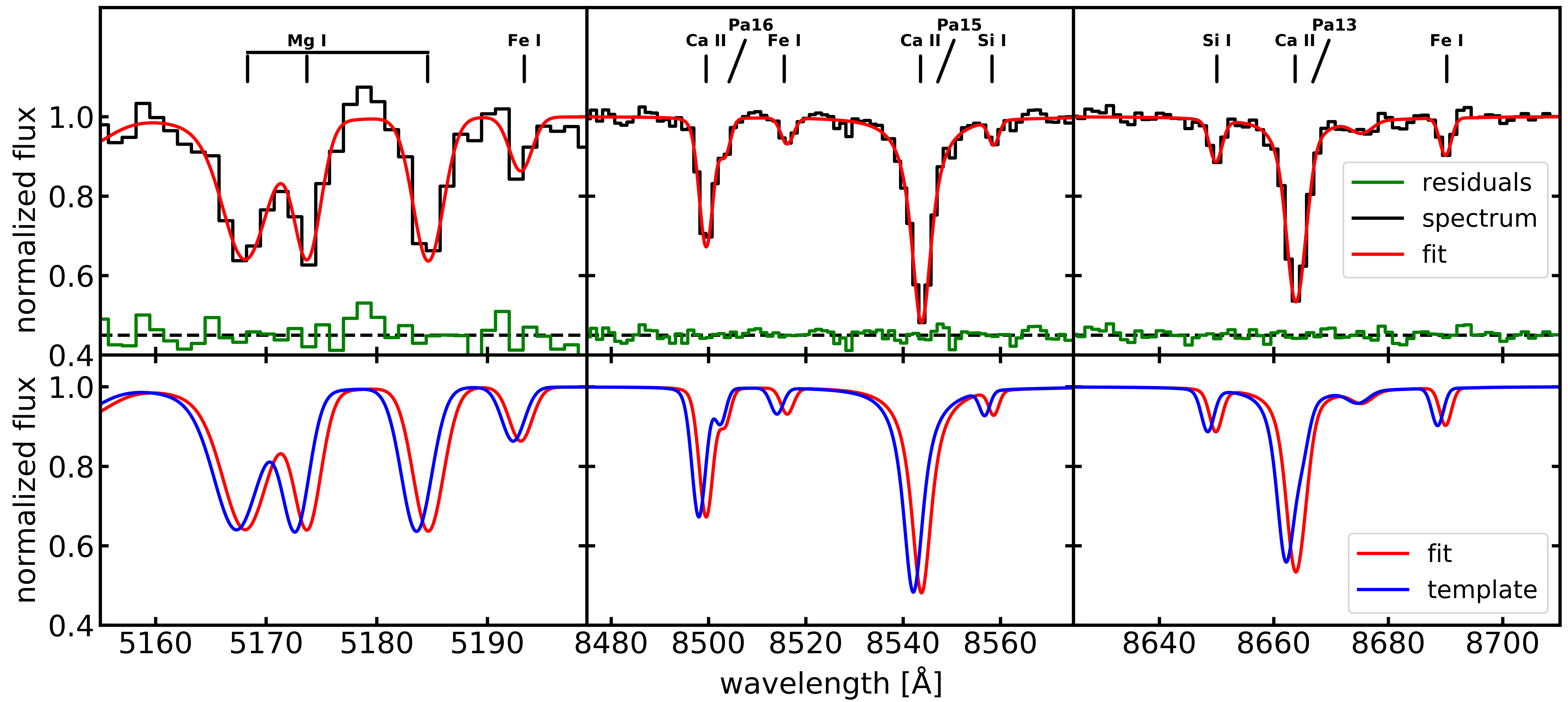}
	\caption{In this figure we show an example of fits of the \ion{Mg}{1} $\lambda\lambda$5367, 5172, 5183 and \ion{Ca}{2} $\lambda\lambda$8498, 8542, 8662 spectral lines. In the top row the extracted and rectified stellar spectrum is plotted (black) including the fitted spectral lines in red. The green line represents the fit residuals (offset by 0.45 from zero in normalized flux units). Additionally, important stellar absorption lines are marked. In the bottom row we show the fitted spectrum in red and the resulting template at rest-frame wavelengths in blue.}
	\label{fig:spectralfit}
\end{figure*}

In Fig.~\ref{fig:spectralfit} we show an example of the fitted template for the \ion{Mg}{1} $\lambda\lambda$5367, 5172, 5183 and \ion{Ca}{2} $\lambda\lambda$8498, 8542, 8662 spectral lines, similar to what can be plotted by executing \texttt{RV\_spectrum.plot}. In black we show the stellar spectrum and in red the fitted spectral lines (top row). In blue we show the created template (bottom row), which is shifted to the rest-frame wavelengths. The green line represents the fit residuals. The small offset of amplitude between the template and the fit is a result of the wavelength dependent shift, which may change the amplitude of the superposition of the primary line and the blend. This does not influence the location of the  position of the centroid of the primary line. We allow for small amplitude adjustments while fitting the RVs (see Sect.~\ref{sec:RVs} to obtain a more stable cross-correlation.

\subsubsection{Important parameters}
The \texttt{line\_fitting} module accepts multiple input parameters. In this section we introduce the mandatory and most important ones but for a detailed description of all available parameters we refer to \url{https://musepack.readthedocs.io/en/latest/index.html}. 

The two mandatory arguments are a spectral line catalog (format: unique line name, wavelength) and an array of names of the primary lines that shall be fitted.

The \texttt{line\_fitting} module also takes multiple important (but not mandatory) keyword arguments. A ``list with blends" (format: prime name, prime wavelength, blend name, blend wavelength, maximum line ratio) may be provided. The maximum line ratio is the maximum ratio of the primary to the blended line. Setting ``autoadjust" to ``True" allows the wavelength range ``llimits", determining the minimum and maximum allowed central wavelength of the fitted lines, to be adjusted during execution depending on the last fitted centroid of the primary line. For all other lines the wavelength shift ($\Delta \lambda / \lambda$) is taken into account accordingly. If ``fwhm\_block" is set to ``True", it limits the fitted line width such that it is never smaller than the dispersion of the instrument to prevent unreasonable and impossible narrow lines. ``adjust\_preference" sets the order in which ``niter" and ``contorder" are adjusted in the cases that either the fit does not converge after ``niter" iterations or it diverges. If set to ``contorder", the polynomial order used to fit the continuum is increased in steps of one (to a maximum of ``max\_contorder"). If the maximum continuum order is reached the wavelength range of the fit (as set in the input catalog) is changed by $\pm5$\,{\AA} until the maximum number of adjustments (``max\_ladjust") is reached. If ``adjust\_preference" is set to ``wavelength" the order of adjustments is reversed. If both parameters reach their set maximum without convergence a warning is issued.

\subsection{Measuring radial velocities}
\label{sec:measure_RV}
After the steps in Sect.~\ref{sec:templates} have been executed the spectral instance also contains the stellar spectrum to be fitted, the template at rest-frame wavelength, as well as a continuum function. By executing \texttt{RV\_spectrum.rv\_fit} the radial velocity fit is initiated. To measure the RVs via cross-correlating the spectrum and the template we use pPXF \citep{Cappellari_04,Cappellari_17}. The \texttt{RV\_spectrum.rv\_fit} instance requires as input the initial guesses in the form [RV, dispersion], similar to pPXF. The dispersion is the difference between the template and the spectrum, which in our case is by definition zero, since the template is created with the fit parameters for each line. We recommend to execute \texttt{RV\_spectrum.rv\_fit\_peak} before \texttt{RV\_spectrum.rv\_fit}, which provides an RV measurement based on the line centroids only, which has proven to be a good initial guess for the RV fit. The RV fit instance executes the following scheme to determine the stellar RVs:

\begin{itemize}
	\item[1] The RV of each primary line is measured. Therefore, pPXF is executed $niter/2$-times (default $niter = 10000$) and for each iteration the uncertainties of the spectrum are randomly reordered. To determine the RV of each line a normal distribution is fit to the RV distribution, while its fitted mean value is considered to be the RV of this line.
	\item[2] If there are three or more absorption lines the median between the RVs and the median absolute deviation (MAD) is determined in order to perform a $3\sigma$ clip. This ensures that lines with ``odd" profiles are removed from the final RV fit. Additionally, only lines with a certain significance are used. This significance is defined by how many standard deviations the spectral line has to be above the local uncertainty (Poisson noise) of the spectrum ($ {\rm amplitude} > {\rm significance} \cdot \sigma$)\footnote{We assume that the local S/N of the local spectrum only changes slowly with wavelength.}. We determined that a significance of 5 is a good value but the user is able to set it as a parameter in the RV fit module.
	\item[3] The final step determines the RV by repeating step 1, using all lines together that were classified as good in the previous step. This time pPXF is executed $niter$-times. At the end a Gaussian is fit to the resulting RV distribution of the $niter$ MC fits. The RV of the star is the mean of this distribution, while the $1\sigma$ standard deviation is defined as the RV uncertainty.
\end{itemize}

Each of the above pPXF runs is independent from the others so these steps can be executed using multi-core processing. The final results of the RV fit is also stored in the spectral object and properly logged.

\subsection{The debug-mode}
The \texttt{RV\_spectrum} class may be executed in debug mode. By activating the debug-mode, all tasks are executed on a single core. The spectral line fit is executed interactively, where each iteration is plotted, all fit parameters are printed to the terminal and the user must confirm before continuing to the next iteration. This mode is useful to follow the individual steps of spectral line fitting (e.g., if the fit keeps diverging) or to trace back error messages, since this can be tricky when running Python on multiple cores.

\subsection{The accuracy and reliability of the RV measurements}
\label{sec:accuracy}
To show that our new method of fitting RVs is reliable and it is indeed possible to determine RVs to an accuracy of 1--$2\,{\rm km}/{\rm s}$, we performed multiple tests and sanity checks. We checked many of the spectral fits by eye to determine the goodness-of-fit and whether weak lines or lines where the spectral fit did not optimally converge where properly discarded.

\begin{figure*}[htb]
	\plotone{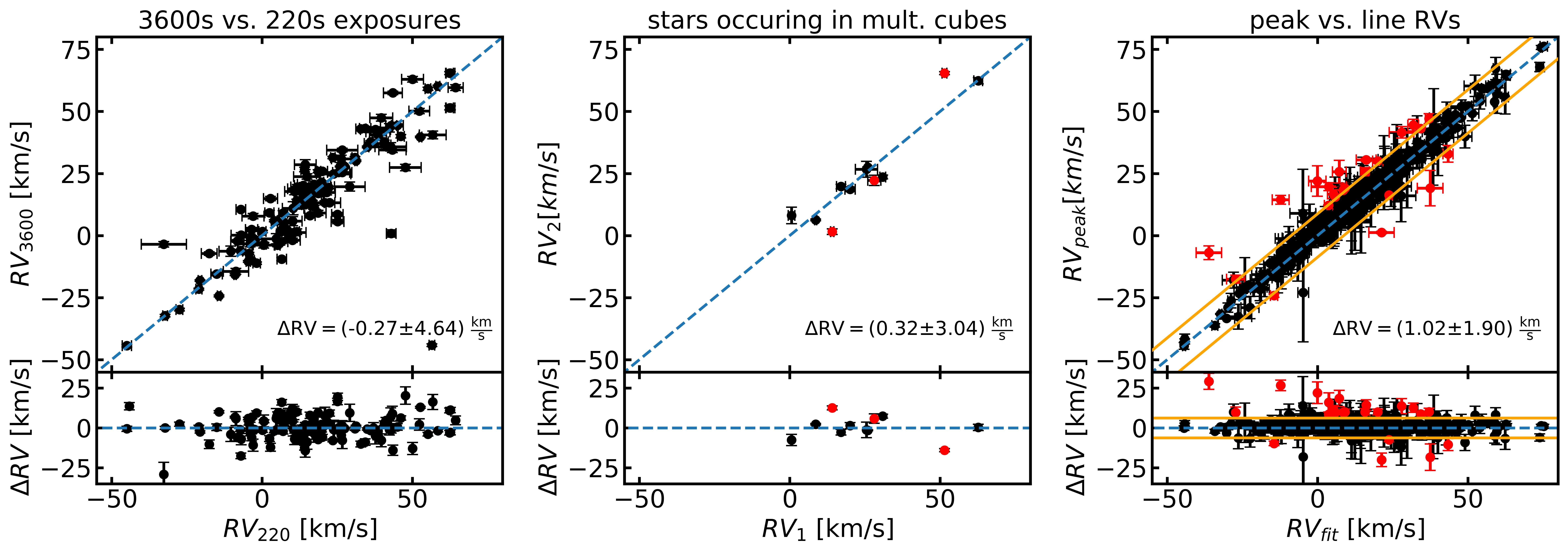}
	\caption{The accuracy of the RV measurements. \textbf{Left panel:} We show all stars that are detected in both the 220\,s and the 3600\,s exposures ($\Delta {\rm RV} = \left(-0.27\pm4.64\right)\,{\rm km}\,{\rm s }^{-1}$). \textbf{Center panel:} We show all stars that are detected in multiple cubes, where the tiles of the mosaic overlap ($\Delta {\rm RV} = \left(0.32\pm3.04\right)\,{\rm km}\,{\rm s }^{-1}$). \textbf{Right panel:} We show the peak vs. the fitted RV measurements ($\Delta {\rm RV} = \left(1.02\pm1.90\right)\,{\rm km}\,{\rm s }^{-1}$). The orange lines represent the $3\sigma$ peak-RV--fit-RV difference used as an additional selection criterion. All stars marked in red do not fulfill this criterion.}
	\label{fig:RV_test}
\end{figure*}

In Sect.~5.1 of \citet{Zeidler_18} we already showed that the RVs determined with \dataset[MUSEpack]{\doi{10.5281/zenodo.3433996}} agree well with the existing literature values for 8 stars ($\Delta {\rm RV} = \left(3.7\pm4.08\right)\,{\rm km}\,{\rm s }^{-1}$). Furthermore, all of the long exposures spatially fully overlap with the short exposures. There is a magnitude range in which stars are faint enough not to saturate in the long exposures but they are still luminous enough so that the extracted spectra from the short exposures have a high-enough S/N to properly measure the stellar RV. 137 stars fulfill this criterion and the median RV offset between the long and short exposures is $\left(-0.27\pm4.64\right)\,{\rm km}\,{\rm s }^{-1}$ (see left frame of Fig.~\ref{fig:RV_test}). Next, we used stars that are detected in the overlap regions of different data cubes. In total 17 stars are observed in multiple cubes and, even though all of them are at the edges of the IFU where the wavelength calibration is less reliable, they agree very well with each other (median offset: $\left(0.32\pm3.04\right)\,{\rm km}\,{\rm s }^{-1}$, see middle frame of Fig.~\ref{fig:RV_test}). Last but not least we checked the correlation between the RV determined by the centroid only and the full RV fit. Even though the uncertainties of the RVs determined with the line centroids are large their median offset is $\left(1.02\pm1.90\right)\,{\rm km}\,{\rm s }^{-1}$ (see right frame of Fig.~\ref{fig:RV_test}). We tested if there is any correlation between the RVs and the absorption line species used to derive them. In order to do so, we used the same set of stars and determined the RVs using different sets of lines (\ion{Ca}{2}, \ion{Mg}{1}, and for the short exposures \ion{Na}{2}) and within the uncertainties the RVs are in agreement. 

For some stars the deviation between the RV measured with the line centroids only and the full RV fit is quite large. After checking the spectral and RV fits we determined that the MC results jumped between two distinct values. Therefore, we decided to use this as an additional quality criterion to create the RV catalog. We flagged all stars whose peak-RV--fit-RV difference (including error bars) is larger than $3\sigma$ (orange lines in the right frame of Fig.~\ref{fig:RV_test}). 30 out of 416 stars or 7.2\% are affected (red stars in the right frame of Fig.~\ref{fig:RV_test}). This additional criterion can only be applied if a statistically significant number of stars is available, but for a small number of stars it is feasible (and recommended) to manually check the results.

The peak-RV--fit-RV difference may be a result of binaries, which we currently are unable to resolve and detect. Due to the limited spectral resolution of MUSE, a minimum projected orbital RV of $\sim 50\,{\rm km}{\rm s}^{-1}$ is necessary to resolve double-peak binaries. Our inspections of the flagged spectra have not shown any hints for such sources. In a future work we plan to analyze possible binaries in those regions where sources are detected in multiple datacubes and incorporate an automatic method to flag those potential binaries. Currently, we suggest to manually check those flagged spectra.

Besides the measured uncertainties for each individual source, the above described tests lead to a combined statistical uncertainty of:

\begin{eqnarray}
\sigma = \sqrt{\sigma_{\rm l,s}^2 + \sigma_{\rm mult}^2 + \sigma_{\rm peak}^2} = 1.10\,{\rm km}\,{\rm s}^{-1}.
\label{eq:stat_uncert}
\end{eqnarray}

This shows that measuring RVs to a 1--$2\,{\rm km}\,{\rm s }^{-1}$ level of accuracy without the existence of spectral libraries is indeed feasible with this method.

\begin{figure*}[htb]
	\plotone{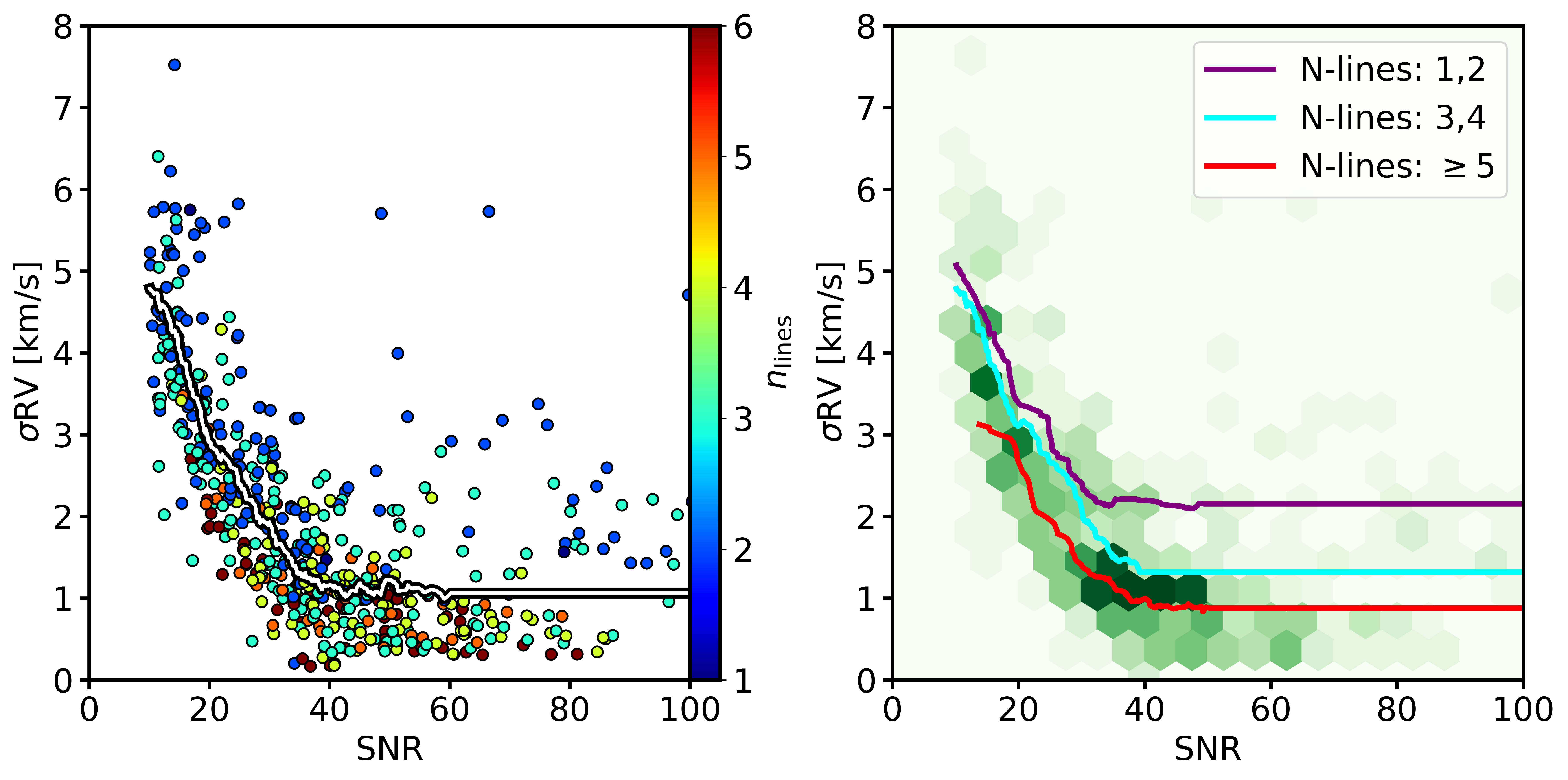}
	\caption{The distribution of the RV uncertainty vs. the S/N ratio of the stellar spectra. \textbf{Left:} Each individual star is color-coded by the number of lines used to perform the final velocity fit. The more lines are used and the higher the S/N of the spectra is, the smaller is the RV uncertainty. The black-white line represents the running mean of the RV uncertainty. \textbf{Right:} A density plot reflecting the $\sigma\left({\rm RV}\right)$-RV plane overplotted with the running mean of the RV uncertainties depending on the used number of lines for the final RV fit: 1--2 lines (purple), 3--4 lines (cyan), or $\geq 5$ lines (red).}
	\label{fig:SNR}
\end{figure*}

We analyzed the correlation between the RV uncertainties, the number of lines used to measure the RVs, and the S/N of the input spectra. In the left panel of Fig.~\ref{fig:SNR} we show the correlation of the RV uncertainties versus S/N of the spectra for each individual star, color-coded with the final (cleaned) number of lines (see Sect.~\ref{sec:measure_RV}) used to measure the final RVs. The black and white line is the running mean of the correlation. In the right panel of Fig.~\ref{fig:SNR} we present a smoothed histogram of the $\sigma\left({\rm RV}\right)$ - S/N relation overplotted with the running mean if 1 or 2 (blue), 3 or 4 (cyan), or $\geq 5$ (red) stellar absorption lines were used to determine the RV. In Tab.~\ref{tab:snr} we list the typical RV uncertainties for a specific S/N and a specific absorption line number used.

Overall, the RV accuracy increases with an increasing S/N and increasing number of used lines, as expected. The accuracy appears to converge for all S/N larger than $\approx 50$ to typical RV accuracies of $0.9\,{\rm km}\,{\rm s}^{-1}$, $1.3\,{\rm km}\,{\rm s}^{-1}$, and $2.2\,{\rm km}\,{\rm s}^{-1}$ with $\geq 5$, 3--4, and 1--2 spectral lines, respectively (see Tab.~\ref{tab:snr}). For a S/N$<10$ the typical RV uncertainty exceeds $5\,{\rm km}\,{\rm s}^{-1}$ and, therefore, we suggest a lower S/N limit of 10 for RV measurements.

\begin{deluxetable}{crrrrr}[htb]
	\tablecaption{The RV uncertainty \label{tab:snr}}
	\tablecolumns{6}
	\tablehead{
		\colhead{n lines} & \multicolumn{5}{c}{S/N}\\
		    & \colhead{10} & \colhead{20} & \colhead{30}& \colhead{40}& \colhead{$\ge 50$}
	}
	\startdata
	all & 4.78 & 2.92 & 1.93 & 1.16 & 1.06 \\
	1 \& 2 & 5.07 & 3.40 & 2.40 & 2.19 & 2.15 \\
	3 \& 4 & 4.78 & 3.10 & 2.10 & 1.32 & 1.30 \\
	$\geq 5$ & --- & 2.65 & 1.41 & 1.00 & 0.88 \\
	\enddata
	\tablecomments{The RV uncertainty dependency on the S/N of the used spectra per number of lines used. All RV uncertainties are given in ${\rm km}\,{\rm s}^{-1}$. For $\geq 5$ lines not enough data were available to obtain a statistically significant RV uncertainty value for a S/N of 10.}
\end{deluxetable}

\section{The W\lowercase{d}2 MUSE radial velocity catalog}
\label{sec:catalog}
As described in the previous section, we measured the RVs for both the long and the short exposures for all stars with a S/N$>10$ for which we extracted clean spectra (see Sect.~\ref{sec:background}). We merged the catalogs from the long and the short exposures and in the case that there are multiple RV measurements for the same source, we will use the RV measurement for which the S/N of the spectrum was the highest. We keep the individual measurements, which we will use in a future work to look for RV shifts caused by binary systems, and we will cross-correlate them with the HST multi-epoch eclipsing binary study by \citet[][in prep.]{Sabbi_19}. The current final catalog\footnote{This is an on-going project and the catalog is subject to future updates, especially after attempting to improve the local stellar background subtraction.} comprises 388 individual stars, of which 117 are cluster members and 271 are field stars, based on the photometric selection using various color-magnitude diagrams \citep{Zeidler_15}. 

\begin{figure*}[htb]
	\plotone{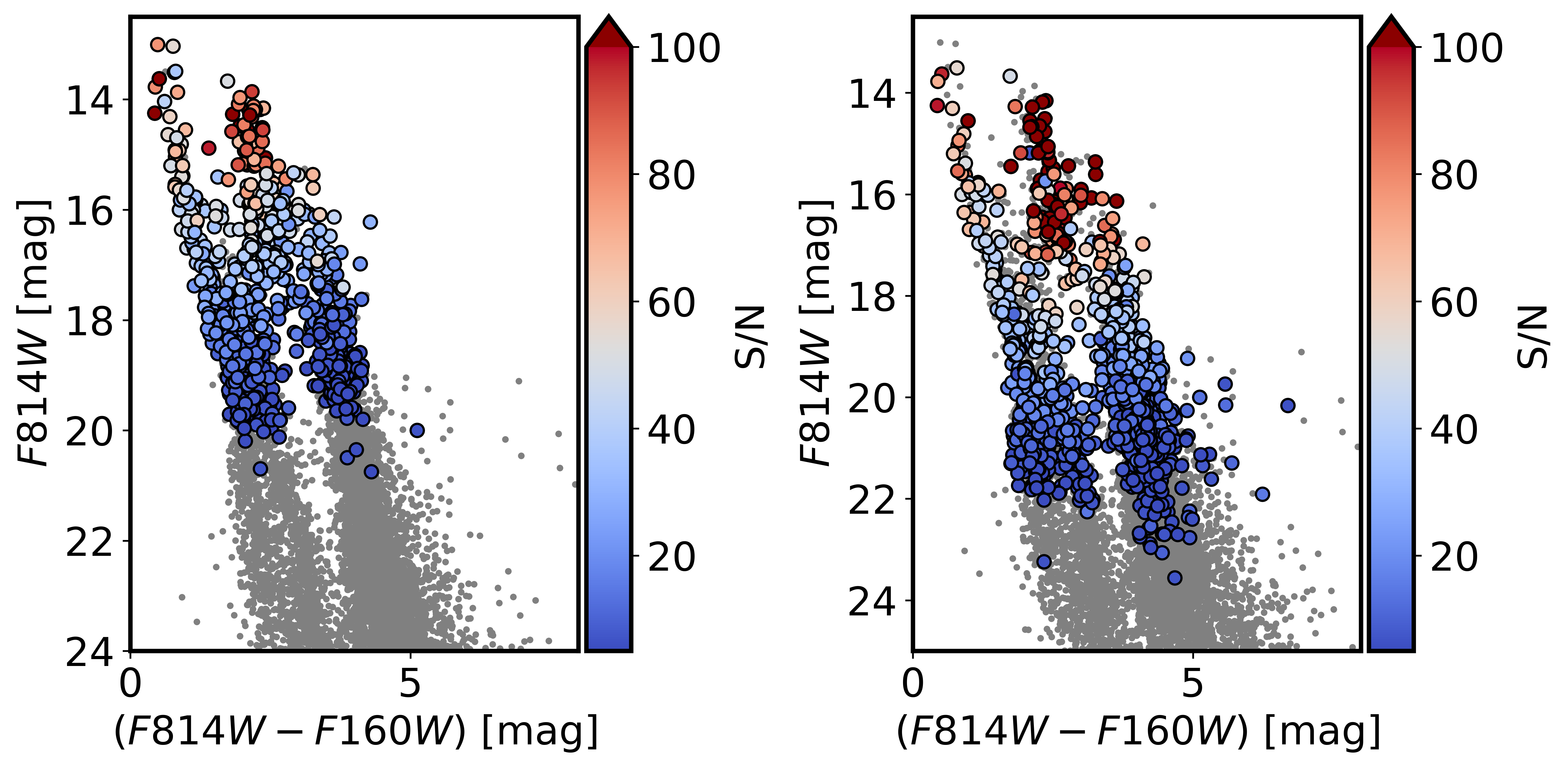}
	\caption{The sources detected with MUSE overplotted on the HST $F814W$ vs. $F814W-F160W$ CMD (gray dots) for the short (left panel) and long (right panel) exposures. The color represents the S/N of the spectra with a minimum S/N of 5.}
	\label{fig:CMD_SNR}
\end{figure*}

In Fig.~\ref{fig:CMD_SNR} we show the $F814W$ vs. $F814W-F160W$ CMD based on the HST photometry (grey points). We marked all stars that are detected in the MUSE dataset color-coded with their S/N. The left panel of Fig.~\ref{fig:CMD_SNR} shows the stars extracted from the short exposures while the right panel shows the stars extracted from the long exposures. Some of the most luminous stars are not detected in the long exposures, either because the MUSE data saturated, or they were outside the area covered by the long exposures (especially true for the cluster center). A S/N$>5$, which we consider as detection threshold, is reached for $F814W = 19.39$\,mag in the short exposures and $F814W = 20.49$\,mag in the long exposures corresponding to a stellar mass of $1.9\,{\rm M}_\odot$ and $1.0\,{\rm M}_\odot$, respectively. A S/N$>10$, which we have demonstrated (see Sect.~\ref{sec:accuracy}) is a lower limit for the reliable measurement of stellar RVs, is reached for $F814W = 18.62$\,mag in short exposures and $F814W = 20.02$\,mag in the long ones or stellar masses of $2.4\,{\rm M}_\odot$ and $1.9\,{\rm M}_\odot$, respectively. All stellar masses are based on a 1\,Myr isochrone at a distance of 4.16\,kpc with a mean visual extinction of $A_V = 6.12$\,mag \citep{Zeidler_15}. A more detailed overview over the brightness and mass limits is given in Tab.~\ref{tab:snr_spec}.

\begin{deluxetable}{rrrrrr}[htb]
	\tablecaption{The magnitude limits \label{tab:snr_spec}}
	\tablecolumns{5}
	\tablehead{\\
		 \multicolumn{1}{c}{S/N} & \multicolumn{2}{c}{short} & \multicolumn{2}{c}{long}\\
		 \multicolumn{1}{c}{} &  \multicolumn{1}{c}{$F814W$} &  \multicolumn{1}{c}{mass} &  \multicolumn{1}{c}{$F814W$} &  \multicolumn{1}{c}{mass}\\
		  \multicolumn{1}{c}{} &  \multicolumn{1}{c}{(mag)} &  \multicolumn{1}{c}{$({\rm M}_\odot)$} &  \multicolumn{1}{c}{(mag)} &  \multicolumn{1}{c}{$({\rm M}_\odot)$}
	}
	\startdata
	100 & 14.55 & 12.7 & 16.12 & 10.9  \\
	80 & 15.15 & 9.7 & 17.61 & 8.5  \\
	40 & 16.56 & 3.6 & 18.40 & 4.4  \\
	20 & 17.67 & 3.1 & 19.33 & 3.0  \\
	10 & 18.62 & 2.4 & 20.02 & 1.9  \\
	5 & 19.38 & 1.9 & 20.49 & 1.0  \\
	\enddata
	\tablecomments{The typical magnitude limits for a given S/N of the stellar spectra for both the short and long exposures. The stellar masses are based on a 1\,Myr isochrone at a distance of 4.16\,kpc with a mean visual extinction of ${\rm A}_V = 6.12$\,mag \citep{Zeidler_15}.}
\end{deluxetable}

\section{Summary and conclusions}
\label{sec:summary}

In this work we introduced the VLT/MUSE IFU dataset of Wd2, a Galactic YMC. This paper is meant to showcase \dataset[MUSEpack]{\doi{10.5281/zenodo.3433996}}, our newly developed Python package, whose main purpose is to measure RVs of stars extracted from IFU datasets in absence of stellar spectral template libraries.

We obtained 16 data cubes of MUSE data covering $11\,{\rm arcmin}^2$ with two different exposure times to study the gas, the luminous upper main sequence stars, as well as the fainter pre-main-sequence down to $F814W = 20.49\,{\rm mag}$ ($1.0\,{\rm M}_\odot$, see Tab.~\ref{tab:snr_spec}). The long exposures covering the pre-main-sequence stars are located around the cluster center (see Fig.~\ref{fig:coverage}) to avoid saturation. Using PampelMuse to extract the stellar spectra, we extracted 1602 stars with a S/N$>5$ (see Tab.~\ref{tab:extr_stars}). The internal kinematics of Wd2 will be presented in a future work.

To reduce the data we use the \dataset[MUSEpack]{\doi{10.5281/zenodo.3433996}} class \texttt{MUSEreduce}, which is a Python wrapper for ESO's data reduction pipeline. It is fully written in Python 3 and, therefore, platform independent and also does not require a GUI, making it convenient in order to run remotely even with slower internet connections or on a computer cluster. As part of \texttt{MUSEreduce} we developed a modified sky subtraction to perform proper sky subtraction without a sky field in crowded regions (e.g., resolved, nearby \ion{H}{2} regions or close-by galaxies). The modifications avoid the oversubtraction of spectral emission lines that are part of both the science target and the Earth's atmosphere, visible as telluric lines (see Sect.~\ref{sec:background} and Fig.~\ref{fig:skysub}). We tested \texttt{MUSEreduce} with the WFM-NOAO and WFM-AO data of Wd2, with WFM-NOAO data of the blue compact dwarf galaxy J0291+0721, a faint extended object \citep{James_19}, as well as NFM-AO data of a T-Tauri disk \citep[][in prep.]{Girard_19}, to ensure that it properly supports all currently provided science modes.

In Appendix~\ref{sec:moon} we showed that it is possible to subtract contamination from the Moon caused by executing the observations under ``bright" conditions. By fitting a lunar model to the data it is possible to correct the data cubes to an extent at which the lunar contamination is only minimal. Although there is still some flux contamination left, which has an influence on the absolute flux measurements, stellar RVs can be measured with the same accuracy as under ``dark" conditions.

We showed that it is possible to measure stellar RVs with an accuracy of 1--$2\,{\rm km}/{\rm s}$ based on MUSE data without the need of a spectral template library using the \dataset[MUSEpack]{\doi{10.5281/zenodo.3433996}} class \texttt{RV\_spectrum} (see Sect.~\ref{sec:RVs}). This new method uses strong stellar absorption lines and fits atomic line libraries to small wavelength ranges around the stellar absorption lines. The fit is performed iteratively by automatically adopting updated input parameters based on the previous iteration within certain user based limits. After convergence is reached a template spectrum is created based on the fitted line parameters and the rest-frame wavelengths of the spectral lines. In a subsequent step, the template is repeatedly cross-correlated against the input spectrum using a MC approach to determine the stellar RV and its uncertainty. Due to this MC approach RV accuracies of $\sim2\,{\rm km}/{\rm s}$ can be reached for the majority of stars, despite the limited spectral resolution of MUSE. The statistical uncertainty is $1.10\,{\rm km}\,{\rm s}^{-1}$. Although S/N dependent, the RV accuracy converges for all S/N larger than $\approx 50$ with typical RV accuracies of $0.9\,{\rm km}\,{\rm s}^{-1}$, $1.1\,{\rm km}\,{\rm s}^{-1}$, and $2.2\,{\rm km}\,{\rm s}^{-1}$ with $\geq 5$, 3--4, and 1--2 spectral lines, respectively. For a S/N$<10$ the typical RV uncertainty exceeds $5\,{\rm km}\,{\rm s}^{-1}$ and, therefore, we suggest a lower S/N limit of 10 for RV measurements.

With the presented method it is now possible to efficiently measure stellar RVs in YMCs to determine their RV profile and dispersion, a key property to understand their evolution in the first few Myr. We will present the results of the RV study for Wd2 in a future paper.

\dataset[MUSEpack]{\doi{10.5281/zenodo.3433996}} is made available for download on Github \url{https://github.com/pzeidler89/MUSEpack.git}. A detailed manual of the main Classes and all of the additional side-modules can be found on \url{https://musepack.readthedocs.io/en/latest/index.html}. More modules will be subsequently added in the future, such as an enhanced local background subtraction for stars in crowded fields.

\acknowledgments
We thank the anonymous referee for their helpful comments and suggestions to improve the quality of the paper.
We thank B. James for providing the dataset J0291+0721 (096.B-0212(A)) and her patience for being the first user of the \dataset[MUSEpack]{\doi{10.5281/zenodo.3433996}} and, especially, \texttt{MUSEreduce}, which highly improved the user-friendliness of the package and tracing errors in the code.
We thank J. Girard for providing the NFM-AO dataset (60.A-9482(A), P.I.: J. Girard), which allowed us to also test this mode.
We thank N. L{\"u}tzgendorf for helping us developing the Monte Carlo module \texttt{MUSEpack.ppxf\_MC}.  
We thank L. Coccato for his patience and help for solving the issues with the Lunar contamination.

P.Z. acknowledges support by the Forschungsstipendium (ZE 1159/1-1) of the German Research Foundation, particularly via the project 398719443.

E.K.G. and A.P. acknowledge support by Sonderforschungsbereich 881 (SFB 881, ``The Milky Way System") of the German Research Foundation, particularly via subproject B05.

These observations are associated with program \#14807. Support for program \#14807 was provided by NASA through a grant from the Space Telescope Science Institute. This work is based on observations obtained with the NASA/ESA \textit{Hubble} Space Telescope, at the Space Telescope Science Institute, which is operated by the Association of Universities for Research in Astronomy, Inc., under NASA contract NAS 5-26555.

Based on observations collected at the European Southern Observatory under ESO programmes 096.B-0212(A), 097.C-0044(A), 099.C-0248(A), 60.A-9482(A).

This work has made use of data from the European Space Agency (ESA) mission {\it Gaia} (\url{https://www.cosmos.esa.int/gaia}), processed by the {\it Gaia} Data Processing and Analysis Consortium (DPAC, \url{https://www.cosmos.esa.int/web/gaia/dpac/consortium}). Funding for the DPAC has been provided by national institutions, in particular the institutions participating in the {\it Gaia} Multilateral Agreement.

\software{PampleMuse \citep{Kamann_13}, ESORex \citep{Freudling_13}, pyspeckit \citep{Ginsburg_11}, MUSE pipeline \citep[v.2.2.0][]{Weilbacher_12,Weilbacher_14}, Astropy \citep{Astropy_18}, Matplotlib \citep{Hunter_07}, pPXF \citep{Cappellari_04,Cappellari_17}, CataPack (Paolo Montegriffo)}

\facilities{VLT:Yepun (MUSE), HST(WFC3,ACS), Gaia(Space)}

\newpage
\appendix
\restartappendixnumbering

\section{Observing under ``bright" conditions}
\label{sec:moon}
During the quality checks we discovered a Moir\'{e}-like pattern in the southern-most long exposure (LONG\_1). This pointing was observed during two different nights (see Tab.~\ref{tab:dataoverview}). LONG\_1a and LONG\_1c were obtained with Moon condition ``bright" but exposure LONG\_1b was observed with Moon condition ``dark". A detailed analysis of the individual cubes and consultation with ESO showed that, even with a Moon distance of $67^{\circ}.83$, observing during ``bright" conditions causes lunar contamination \citep[e.g.,][]{Patat_04}. This contamination is an additive component in the spectrum, which is not properly handled by the MUSE data reduction pipeline.

\subsection{Removing the lunar spectrum}
To correct for the lunar contamination, we attempted to subtract the lunar component from the individual data cubes LONG\_1a and LONG\_1c. The Lunar spectrum observed with ground telescopes is in principle a Solar spectrum reflected by the Moon and transmitted through the Earth's atmosphere. The transmission efficiency of the Earth's atmosphere depends on wavelength, the observing conditions (temperature, wind, and observation site),  the position of the Moon on the sky, the Moon phase (influences the albedo), as well as the Moon angle (the angular separation between the Moon and the observed target) \citep[for more details we refer to][]{Patat_04}. This means that it is very difficult to model the lunar component for a specific observation \citep[e.g.,][]{Schaefer_91}. To estimate the sky contribution ESO provides an online tool ``SkyCalc" (Sky Model Calculator\footnote{\url{https://www.eso.org/observing/etc/bin/gen/form?INS.MODE=swspectr+INS.NAME=SKYCALC}}), with which it is possible to estimate the flux contribution of the sky for any specific observing conditions. We used the lunar model for the observing conditions of LONG\_1a and LONG\_1c and fitted this model together with an extracted, high S/N sky spectrum of the LONG\_1b cube. We used pPXF, which also allows for small RV corrections between individual templates, to find the best matching linear combination of these two spectra. To better match the continuum we added a fourth-order Legendre-polynomial series. We then combined the weighted lunar spectrum and the Legendre-polynomial and added it as a continuum spectrum to the MUSE data reduction process. This ensures that the continuum spectrum gets properly convolved with the local LSF.

\subsection{Quality checks of the correction}
\begin{figure*}[htb]
	\plotone{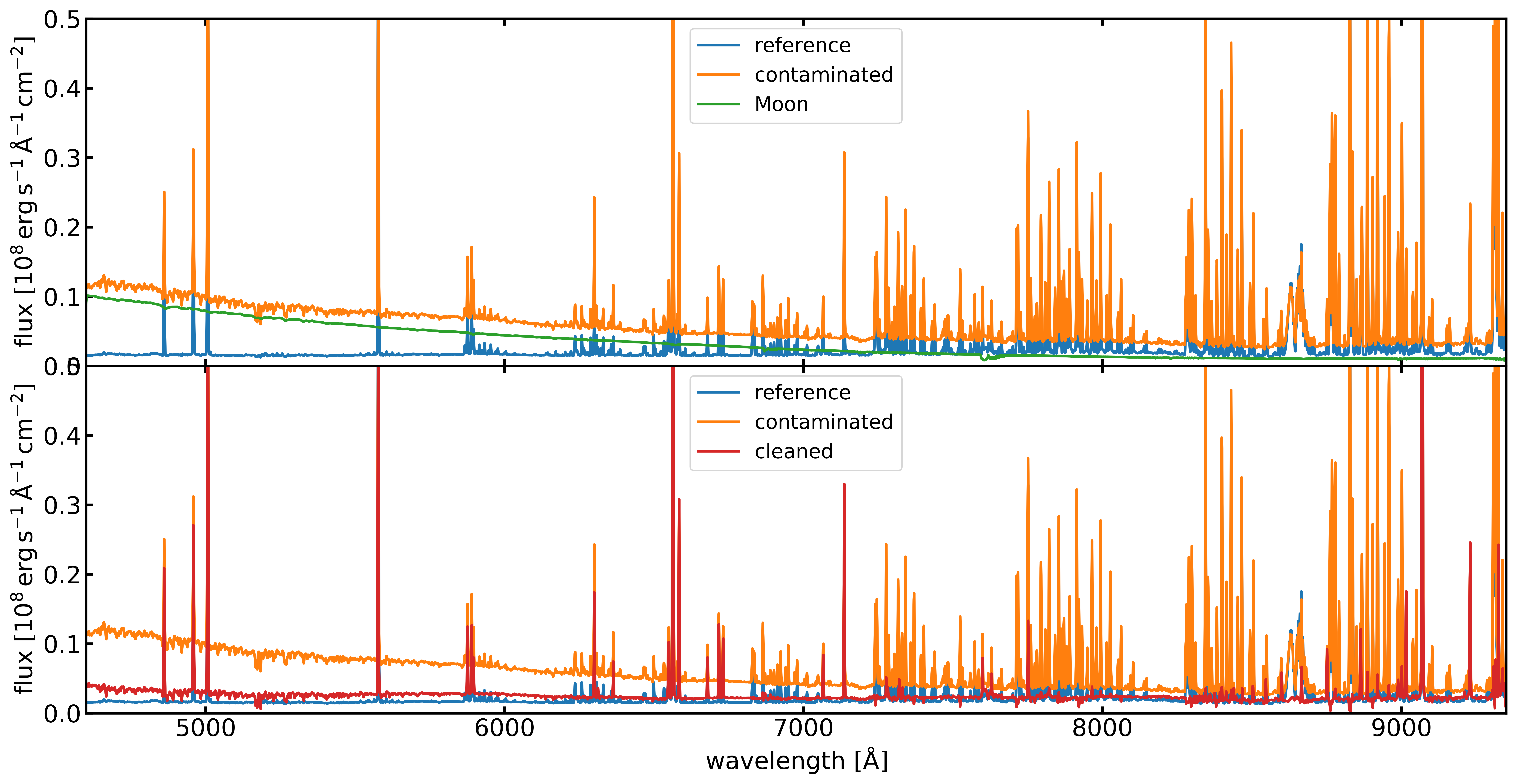}
	\caption{We show the removal of the lunar contamination from cube LONG\_1a (LONG\_1c is analogues). \textbf{Top:} The spatially collapsed spectrum of the contaminated data cube (orange) and the clean cube (LONG\_1b) used as reference (blue). The stars were masked in both cubes to avoid a bias toward absorption lines. The green spectrum is the best fitting lunar spectrum. \textbf{Bottom:} The spatially collapsed spectrum of the contaminated data cube (orange) and the clean cube (LONG\_1b) used as reference (blue). The stars were masked in both cubes to avoid a bias toward absorption lines. In red we show the spectrum of cube LONG\_1a after applying the correction using the ESO data reduction pipeline.}
	\label{fig:lunar_correction}
\end{figure*}

In Fig.~\ref{fig:lunar_correction} we show the spatially collapsed spectrum of the contaminated data cube (orange) and the clean cube (LONG\_1b) used as reference (blue) as well as the lunar spectrum (green spectrum, top panel) and the spectrum after applying the correction using the ESO data reduction pipeline (red spectrum, bottom panel). Fig.~\ref{fig:lunar_correction} shows the spectra for data cube LONG\_1a. The correction for LONG\_1c is analogous. Although this procedure does not fully remove the lunar component (see bottom panel of Fig.~\ref{fig:lunar_correction}) it significantly improves the reduced data allowing the MUSE data reduction pipeline to properly combine the three dither positions.

\begin{figure}[htb]
	\epsscale{0.5}
	\plotone{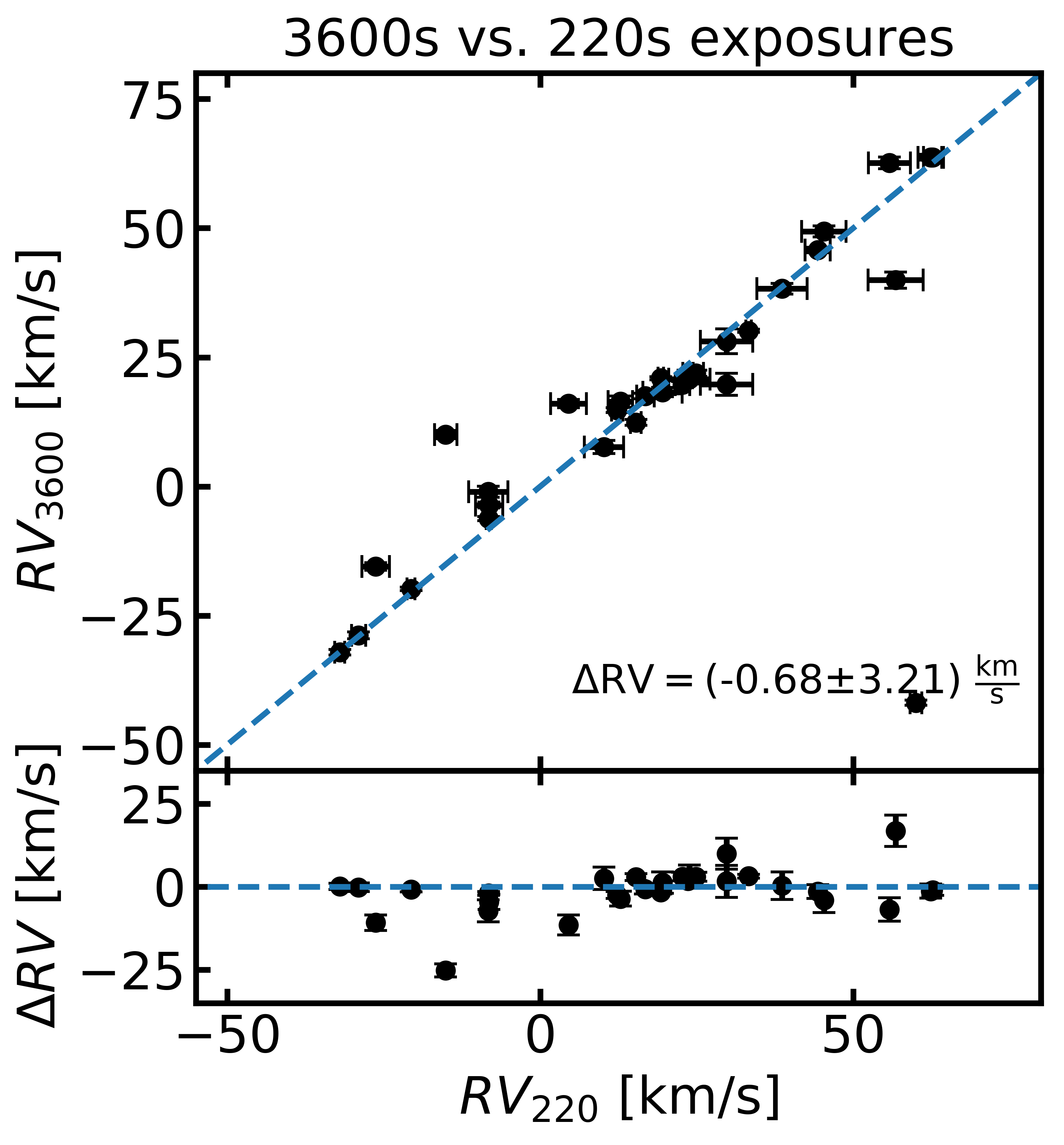}
	\caption{The stellar RV comparison between the long (LONG\_1) and short (SHORT-MID\_1) exposure to determine the reliability of the data cube after the removal of the lunar spectrum.}
	\label{fig:lunar_RV}
\end{figure}

We compared the stellar RVs of the long (LONG\_1) and short (SHORT-MID\_1) exposure (for a detailed description on how the RVs are determined, we refer to Sect.~\ref{sec:RVs}) to determine whether the lunar residuals influence the RV measurements. In Fig.~\ref{fig:lunar_RV} we show the RVs of the stars detected in both data cubes. The mean offset between the two exposure times is $\Delta {\rm RV} = \left(-0.68\pm3.21\right)\,{\rm km}\,{\rm s }^{-1}$, which agrees very well with the full sample ($\Delta {\rm RV} = \left(0.55\pm3.04\right)\,{\rm km}\,{\rm s }^{-1}$, see left panel of Fig.~\ref{fig:RV_test}).

\bibliographystyle{aasjournal}
\bibliography{../../bibliography/Wd2_bibliography}

\end{document}